\documentclass[12pt,a4paper]{article}
\pdfoutput=1
\usepackage[bottom]{footmisc}
\usepackage{graphicx}
\usepackage{appendix}
\usepackage{amsmath,amsfonts,amssymb,setspace}
\usepackage{braket,xcolor}
\usepackage{caption}
\usepackage{subcaption}
\usepackage{hyperref}
\usepackage[capitalize]{cleveref}
\usepackage{cite}
\usepackage{tocloft}
\usepackage{comment}
\usepackage{soul}
\usepackage[margin=0.75in]{geometry}
\hypersetup{colorlinks=false, linkcolor=blue, citecolor=red}
	\def\be{\begin{equation}}
	\def\ee{\end{equation}}
	
	\DeclareMathOperator{\Tr}{Tr}

\newcommand{\p}{\prime\,}
	\newcommand{\vs}[1]{\vspace{#1 mm}}
	
\begin{document}
	\begin{flushright}
		%
	\end{flushright}
	\begin{center}
		{\Large{\bf Decoherence, Entanglement Negativity and Circuit Complexity for Open Quantum System}}\\

\vs{10}

{\large
Arpan Bhattacharyya${}^{a,\,}$\footnote{\url{abhattacharyya@iitgn.ac.in
}},Tanvir Hanif ${}^{b,\,}$\footnote{\url{thanif@du.ac.bd}}, S. Shajidul Haque${}^{c,d,\,}$\footnote{\url{shajid.haque@uct.ac.za}},\\ Arpon Paul ${}^{e,\,}$\footnote{\url{paul1228@umn.edu}}}

\vskip 0.3in

{\it ${}^{a}$ Indian Institute of Technology, Gandhinagar, Gujarat-382355, India}\vskip .5mm

{\it ${}^{b}$ Department of Theoretical Physics, University of Dhaka, Dhaka-1000, Bangladesh}\vskip .5mm

{\it ${}^{c}$
High Energy Physics, Cosmology \& Astrophysics Theory Group \\ and \\ The Laboratory for Quantum Gravity \& Strings \\ Department of Mathematics and Applied Mathematics, \\ University of Cape Town, Cape Town-7700, South Africa }\vskip .5mm

{\it ${}^{d}$ National Institute for Theoretical and Computational Sciences (NITheCS)\\ South Africa}

{\it ${}^{e}$ University of Minnesota Twin Cities, Minneapolis, Minnesota 55455, USA}

\vskip.5mm

\end{center}

\vskip 0.35in

\begin{abstract}
In this paper, we compare the saturation time scales for complexity, linear entropy and entanglement negativity for two open quantum systems. Our first model is a coupled harmonic oscillator, where we treat one of the oscillators as the bath. The second one is a type of Caldeira Leggett model, where we consider a one-dimensional free scalar field as the bath. Using these open quantum systems, we discovered that both the complexity of purification and the complexity from operator state mapping is always saturated for a completely mixed state. More explicitly, the saturation time scale for both types of complexity is smaller than the saturation time scale for linear entropy.
On top of this, we found that the saturation time scale for linear entropy and entanglement negativity is of the same order for the Caldeira Leggett model. 
\end{abstract}

	\newpage

\tableofcontents






\section{Introduction}
\label{sec:intro}

Understanding the transition of quantum to classical behaviour is an important problem in many branches of physics, such as in low-temperature physics \cite{horodecki2013fundamental, Skrzypczyk_2014}, early universe cosmology \cite{Lesgourgues:1996jc}, quantum computation \cite{PhysRevA.52.R2493,Aharonov_2000} etc. The first step towards this transition is believed to be decoherence \cite{Habib:1998ai}, which means loss of quantum coherence that originates from the interaction of a quantum system with its surrounding environment.  For a comprehensive review of this subject, interested readers are referred to \cite{Schlosshauer:2019ewh, Zurek:2003zz}. The decoherence or the amount of mixedness for an open quantum system \footnote{For a comprehensive review of recent developments towards understanding the physics of open quantum systems, interested readers are referred to \cite{open}.} can be quantified by a quantity known as the linear entropy \cite{Zurek:2003zz,Peters_2004}. \par

Apart from the amount of mixedness, quantum entanglement is also a crucial feature of a quantum state. For a closed system, von Neumann entropy \cite{Horodecki_2009}, is a useful measure. But von Neumann entropy is not always a useful measure to quantify quantum correlation for an open system since it also captures the classical correlations. For an open quantum system, the entanglement negativity is a useful quantity that measures the quantum correlation between the system and its environment. In \cite{Peres:1996dw, Horodecki:1996nc, Eisert:1998pz, Vidal:2002zz, Plenio:2005cwa} the negativity was proposed as a measure of quantum entanglement for mixed states. Entanglement negativity, which stems from the criteria for separability of mixed state, is computed by taking the trace norm of the partial transpose of the density matrix. In recent times, the interplay between mixedness and quantum correlation between has been subject to lots of study \cite{Verstraete_2001,PhysRevA.62.022310,PhysRevA.67.022110,PhysRevA.64.012316,PhysRevLett.92.087901,Adesso:2004hs, Benatti:2006pw,Singh_2015,delCampo:2019qdx} \footnote{These references are not exhaustive by no means. Interested readers are encouraged to consult the references and citations of these references.}.


Another quantity that got much attention in recent years to characterize various aspects of quantum systems is the \textit{circuit complexity}. Although the original interest came from AdS/CFT in some black hole settings \cite{Susskind:2014moa,Susskind:2014rva,Brown:2015bva,Brown:2015lvg,Carmi:2016wjl} as an extension to the entanglement entropy, it was already quite well known in quantum information theory and is now widely used for probing various quantum systems \cite{Jefferson,Chapman:2017rqy,Bhattacharyya:2018wym,Caputa:2017yrh,me1,Bhattacharyya:2018bbv,Hackl:2018ptj,Khan:2018rzm,Camargo:2018eof,Ali:2018aon,Caputa:2018kdj,Guo:2018kzl,Bhattacharyya:2019kvj,Flory:2020eot,Erdmenger:2020sup,Ali:2019zcj,Bhattacharyya:2019txx,cosmology1,cosmology2,DiGiulio:2020hlz,Caceres:2019pgf,Bhattacharyya:2020art,Liu_2020,Susskind:2020gnl,Chen:2020nlj,Czech:2017ryf,Chapman:2018hou,Chapman:2019clq,Doroudiani:2019llj,Geng:2019yxo,Guo:2020dsi,Haque:2021hyw,Haque:2021kdm,Caputa:2022yju,Caputa:2022eye,Couch:2021wsm,Erdmenger:2021wzc,Chagnet:2021uvi,Koch:2021tvp,Bhattacharyya:2022ren,Banerjee:2022ime,Bhattacharya:2022wlp} \footnote{This list is by no means exhaustive. Interested readers are referred to these reviews \cite{Chapman:2021jbh, Bhattacharyya:2021cwf}.}. The idea of circuit complexity comes from the theory of quantum computation. It is based on quantifying the minimal number of operations or gates required to build a circuit that will take one from a given reference state ($|\psi_R\rangle$) to the desired target state ($|\psi_T\rangle$). In this paper, we will follow the approach pioneered by Nielsen \cite{NL1,NL2,NL3} to compute the circuit complexity.\par

In \cite{open2,PhysRevD.105.046011}, it has been shown that the complexity can be a useful probe of open quantum systems. In this paper, we further explore whether the circuit complexity can be a useful indicator to estimate if the system is in a mixed or pure state. This is an important question from the perspective of the open quantum system and complements the analysis done in \cite{open2,PhysRevD.105.046011}. We investigate the time evolution of linear entropy and negativity along with complexity for several representative systems. We find some general features of the behaviour of the linear entropy, negativity and complexity in different systems. We have considered two particular solvable models of a system coupled to a  bath with several parameters. First, we consider two harmonic oscillators, where we treat one of the oscillators as the system and the other as the environment/bath. Secondly, we study a harmonic oscillator coupled to a one-dimensional bosonic (free) field theory. The target states are generated by quench in both models. Specifically, we start with the system and bath decoupled and we suddenly turn on the coupling between the system and the bath and also change the parameters in the Hamiltonian abruptly. We take the resulting state, trace out the bath, and scrutinize the reduced density matrix of the remaining oscillator --- we show results (as a function of time) for the complexity and the linear entropy, which is a measure of the purity of the system. Furthermore, we investigate the quantum correlation between the system and bath using negativity by considering the total density matrix of the combined system and bath. We found that entanglement negativity negativity saturates only for the Caldeira Leggett model and the saturation time scale of linear entropy and negativity coincides. Moreover, the circuit complexity always saturates when the system becomes completely mixed. 

The paper is organized as follows. In Section~\ref{EntropyDef} and \ref{ComDef} we provide a brief review of linear entropy, entanglement negativity and the complexity of purification, complexity by operator-state mapping, respectively. Section~\ref{OQSDef} provides the details of the two open quantum systems that we have considered. It also includes the main results and the findings. Finally, we conclude with a discussion and future direction in Section~\ref{discussion}. Some useful details regarding computation of circuit complexity and correlation function for the Caldeira Leggett model are given in the Appendix~\ref{d},\ref{A}, \ref{B} and \ref{C}.

\section{Linear Entropy and Entanglement Negativity }\label{EntropyDef}
Consider the system as a combination of two subsystems $A$ and $B$ (for example, the harmonic oscillator, and the bath); one considers the reduced density matrix of subsystem-$A$, upon tracing out subsystem-$B$
\begin{equation}
 \hat{\rho}_A = {\rm Tr}_B[ \hat{\rho} ] \ , 
\end{equation}
where $\hat{\rho}$ is the density matrix of the entire system.
A central measure of entanglement is the $\alpha^{\rm th}$-Renyi entropy, $S_{\alpha}$, ($\alpha > 1$) 
is defined by
\begin{subequations}
\begin{equation}
 S_{\alpha} = - \frac{1}{\alpha - 1} 
   \ln {\rm Tr} [ \hat{\rho}_A^{\alpha} ]  \ ; 
\end{equation}
taking $\alpha \rightarrow 1^+$ gives the {\sl entanglement entropy}, $S$,
\begin{equation}
 S = - {\rm Tr} [ \hat{\rho}_A \ln \hat{\rho}_A ] \ .
\end{equation}
\end{subequations}
\subsection*{Entanglement Negativity:}
However, when the system is described by a mixed state then the entanglement entropy is not a good measure of quantum  correlation (e.g $S_A \neq S_B$ in general for mixed states). There are several other measures \cite{Horodecki_2009,bengtsson2017geometry}, one of them is \textit{Entanglement Negativity} \footnote{There are other measures of quantum entanglement in mixed states e.g. Mutual Information, Entanglement of Purification. Interested readers are  referred  to \cite{bengtsson2017geometry} for more details of various correlation measures.}.
This is defined by taking the trace norm of the partial transpose of the density matrix. Consider a bipartite state $\rho_{AB}$, consists of sub-systems A and B. The partial transposition with respect to sub-system B of $\rho_{AB}$ that is expanded in a given local orthonormal basis as $\rho = \sum \rho_{ij,kl}\ket{i}\bra{j} \otimes \ket{k}\bra{l}
$ is defined as
\begin{align}
\rho^{T_B} :=
\sum_{i,j,k,l} \rho_{ij,kl}\ket{i}\bra{j} \otimes \ket{l}\bra{k}
\end{align}
Note that the spectrum of the partial transposition of the density matrix is independent of the choice of the basis and is also independent of whether the partial transposition is taken over sub-system A or B. The positivity of the partial transpose of a state is a necessary condition for separability of the density matrix \cite{PhysRevLett.77.1413}. This criterion of separability motivates one to consider the entanglement negativity, which is related to the absolute value of the sum of the negative eigenvalues of $\rho^{T_B}$ \cite{PhysRevA.65.032314,doi:10.1080/09500349908231260,10.5555/2011706.2011707}. Explicitly, this is defined as
\begin{align}
\mathcal{N}(\rho):= \frac{1}{2}\left(\|\rho^{T_B}\|-1\right), \label{negativity}
\end{align}
where $\|A\|\equiv \Tr{\sqrt{A^\dagger A}}$, is the trace norm. It can be shown that $\mathcal{N}(\rho)$ vanishes  for unentangled states, i.e, for separable density matrices $\rho$. Another relevant quantity named as the \textit{Logarithmic Negativity}, $E_{\mathcal{N}}$ is defined as below: 
\begin{align}
E_{\mathcal{N}}(\rho)&:= \log_2{\left\|\rho^{T_B}\right\|} \\
&= \log_2{\left(2\mathcal{N}(\rho)-1\right)} .
\end{align}
These quantities have been explored in recent times in the context of field theory and quantum many-body systems \cite{Calabrese:2012ew,Calabrese:2012nk,Calabrese:2013mi,Alba_2013,Calabrese:2014yza,PhysRevB.90.064401,Ruggiero_2016,Ruggieroa_2016,Blondeau_Fournier_2016,Eisler_2014,Coser_2014,Hoogeveen_2015,PhysRevB.92.075109,PhysRevA.88.042319,PhysRevA.88.042318,Wen_2016,Wena_2016,Kudler_Flam_2019,MohammadiMozaffar:2017chk}.\\

\noindent
\textit{Entanglement Negativity for Gaussian State:} In this paper, we are interested in the entanglement negativity for a bipartite mixed Gaussian state. A Gaussian state can be characterized by the first moments of the phase-space variables and the covariance matrix $\sigma$ defined as below:
\begin{align}
\sigma_{ij}= \frac{1}{2}\langle X_i X_j + X_j X_i\rangle - \langle X_i \rangle \langle X_j \rangle \ ,\label{cvdef}
\end{align}
where $X_i'$s denote the phase-space variables that satisfy the canonical commutation relations, $[X_i, X_j] = 2\, i \,\Omega_{ij}$. Here $\Omega_{ij}$ is the symplectic form. In case of bipartite system, the covariance matrix can be constructed out of three $2\times 2$ matrices, $\alpha,\beta, \gamma$, in the following way:
\begin{align}
\sigma = \begin{pmatrix}
\alpha & \gamma \\
\gamma^T &\beta
\end{pmatrix},
\end{align}
where $\alpha$ and $\beta$ corresponds to the correlators among the phase-space variables of subsystem A and B respectively, whereas $\gamma$ refers to the cross-correlators between subsystem A and B. Now, the partial transposition of the bipartite Gaussian density matrix $\rho$ changes the covariance matrix $\sigma$ into a new matrix $\tilde{\sigma}$ where the $det$ of the cross-correlator $\gamma$ flips the sign, i.e, $\det{\gamma} \rightarrow - \det{\gamma}$ \cite{PhysRevA.72.032334,PhysRevLett.84.2726}. The negativity can be defined in terms of the symplectic eigenvalues $\tilde{\nu}_{\pm}$ of the new covariance matrix $\tilde{\sigma}$ in the following way:
\begin{align}
\mathcal{N}(\rho) = \max\left[0,\frac{1-\tilde{\nu}_-}{2\tilde{\nu}_-}\right]. \label{negdef}
\end{align}
Here, symplectic eigenvalues of $\tilde{\sigma}$ can be evaluated as
\begin{align}
\tilde{\nu}_{\pm} = \sqrt{\frac{1}{2}\left(\Delta(\tilde{\sigma})\mp \sqrt{\Delta(\tilde{\sigma})^2 - 4 \det{\tilde{\sigma}}}\right)}
\end{align}
where, 
\begin{align}
\Delta(\tilde{\sigma}) \equiv \det{\alpha}+\det{\beta} - 2 \det{\gamma}\,.
\end{align}

\subsection*{Linear Entropy:}
It is well known that apart from quantum entanglement, the amount of mixedness is also a crucial feature of a quantum state. For any quantum state, interaction with the environment leads to decoherence. In other words, noise coming through these interactions introduces mixedness in the quantum system, leading to the loss of information \cite{Schlosshauer:2019ewh}. Therefore, quantifying (as well as classifying) the amount of quantum correlation and mixedness and the interplay between them is an important question in quantum information theory \cite{Verstraete_2001,PhysRevA.62.022310,PhysRevA.67.022110,PhysRevA.64.012316,PhysRevLett.92.087901,Adesso:2004hs, Benatti:2006pw,Singh_2015,delCampo:2019qdx}. A well known quantity to characterize the degree of decoherence (or mixedness) that the subsystem-$A$
experiences, due to subsystem-$B$ is called purity which is defined follows \cite{Zurek:2003zz,Peters_2004} \footnote{There are other measures of mixedness, e.g. relative entropy \cite{Baumgratz_2014}, skew information \cite{Girolami_2014}. However, for our case, we will mainly focus on linear entropy, given the simplicity of its computation. Also, \cite{Streltsov:2015xia} proposed a family of measures of quantum coherence/decoherence by using measures of entanglement. Interested readers are referred to this reference for more details and the citations of it for more details.}  :
\begin{equation}
 P = {\rm Tr}[\hat{\rho}_A^2]  \ .
\label{puritydefinition}
\end{equation}
Note that $1/d \leq P \leq 1$ with $d$ being the dimension of the Hilbert space ---
$P=1$ for a pure state, while $P=1/d$ for a completely mixed state. From Purity one can define linear entropy as
\begin{equation}
S_l=1-P
\end{equation}
%

In this paper we want to investigate the saturation time scale of linear entropy and entanglement negativity for a mixed state. To facilitate the comparison we will use the normalized entanglement negativity. 

\section{Circuit Complexity for Density Matrix} \label{ComDef}
Our goal in this paper is to provide a proof of principle argument to establish the circuit complexity as a probe of the mixedness of the system. In this section we briefly discuss methods of computing circuit complexity. There are several methods for computing circuit complexity for closed quantum systems (pure states). In the Appendix~\ref{d}, we provide a brief review on circuit complexity. We mainly follow the Nielsen approach \cite{NL1,NL2,NL3, Jefferson}. Computing complexity for open quantum systems is more involved just like the entanglement entropy. In this section we will briefly discuss two different techniques of computing circuit complexity for mixed states \cite {PhysRevD.105.046011, open2}.
\subsection*{Complexity of Purification (COP):}
The mixed state can be purified to a pure state $\ket{\psi}$ in an enlarged Hilbert space $\mathcal{H}_{pure}  \otimes \mathcal{H}_{anc}$, where $\mathcal{H}_{anc}$ corresponds to an \textit{ancillary} set of degrees of freedom. However, one needs to make sure that the trace of the density matrix of $\ket{\psi}$ over the ancillary Hilbert space will give the original mixed state, $\hat{{\rho}}_{\rm mix}$.
\begin{align}\Tr_{\mathcal{H}_{\text{anc}}}\left(\ket{\psi}\bra{\psi}\right) =  \hat{{\rho}}_{\rm mix}
\end{align}
The expectation values of operators acting in the initial Hilbert space $\mathcal{H}$ remain the same
under purification so that observables are preserved by purification.
\begin{align}\langle \hat{{\mathcal{O}}}\rangle = \Tr_{\mathcal{H}}(\hat{{\rho}}_{\rm mix}\hat{\mathcal{O}}) = \Tr_{\mathcal{H}_{anc}}(\bra{\psi}\hat{\mathcal{O}}\ket{\psi})
\end{align}
Because the choice of the ancillary Hilbert space $\mathcal{H}_{\text{anc}}$is \textit{not} unique, the purification process is arbitrary and depends on the choice of ancillary Hilbert space.
For example, we can construct a set of pure states ${\ket{\Psi}_{\alpha,\beta, ...}}$, 
specified by several parameters $\alpha,\beta, ...$ all of which satisfy the purification requirement. We can minimize the quantity of interest (for example, complexity or entanglement entropy) of the purification process with respect to the parameters, by choosing one particular set of parameters. In this work, we intend to study the complexity of the mixed state; therefore, we will find the minimum of the complexity of the set of purifications ${\ket{\Psi}_{\alpha,\beta, ...}}$ of $\hat{\rho}_{\text{mix}}$, obtaining the COP,
\begin{align}
\mathcal{C}_\rho \equiv \underset{\alpha,\beta, ...}{\text{min}} \mathcal{C}\left(\ket{\Psi}_{\alpha,\beta,...}, \ket{\Psi_R}\right)
\end{align}
where we made explicit the dependence of the complexity of the pure state on the reference state $\ket{\Psi_R}$ and computation of it is detailed in Appendix~\ref{d} (in particular please refer to Eq.~\ref{complexity}). In general optimization is quite difficult, but we will make some simplifying assumption. Motivated by the observations from \cite{Caceres:2019pgf,Bhattacharyya:2018sbw, Bhattacharyya:2019tsi}, we will choose a minimal purification such that the size of the original (sub) system is the same as the auxiliary system as well the purified state will be again a Gaussian state.

\subsection*{Complexity by Operator-State Mapping:}
We can compute complexity in a different way by fixing the choice of the ancillary Hilbert space. In this new definition, we purify the reduced density matrix $\hat{\rho}_{mix}$ using the technique of operator-state mapping (also known as channel-state mapping) \cite{JAMIOLKOWSKI1972275, CHOI1975285,PhysRevA.87.022310,Hosur:2015ylk}.
Suppose, we consider an operator in the Hilbert space $\mathcal{H}$ with representation $\mathcal{O} = \sum_{m,n} \mathcal{O}_{mn}\ket{n}\bra{m}$. In the state-operator mapping scheme, we
associate a state $\ket{\mathcal{O}}$ to $\hat{\mathcal{O}}$ by altering the bra to a ket,
\begin{align}
\hat{\mathcal{O}}=\sum_{m,n} \mathcal{O}_{mn}\ket{n}\bra{m} ~~\Leftrightarrow~~ \ket{\mathcal{O}} = \frac{1}{\sqrt{\Tr{\left[\mathcal{O}\right]}}} \sum_{m,n} \mathcal{O}_{mn} \ket{m} \otimes \ket{n}_{\text{anc}}
\label{Eqn:1}
\end{align}
The state $\ket{\mathcal{O}}$ exists in the extended Hilbert space $\mathcal{H}\otimes\mathcal{H}_{\text{anc}}$.
Once again, we have denoted the extra copy of $\mathcal{H}$ as $\mathcal{H}_{\text{anc}}$ to
distinguish it from the original. In that sense, this process of
operator-state mapping is a special case of the purification
discussed above. The most important difference is that the state $\ket{\mathcal{O}}$ in Eq. \Ref{Eqn:1} associated to the operator $\hat{\mathcal{O}}$ is unique, as there are no free parameters introduced in the mapping. Hence, the complexity associated with the
operator-state mapping does not require a minimization. Last but not the least, given this effective wavefunction as mentioned in Eq.~\ref{Eqn:1} we can compute the circuit complexity using Nielsen's method \cite{NL1,NL2,NL3} as detailed in Appendix~\ref{d}. For our case, this wavefunction will be of the Gaussian form as mentioned in Eq.~\ref{wavefunction} and cosequently the compelxity will be given by Eq.~\ref{complexity}.\par
 In the rest of the paper, we will like to investigate the time evolution of complexity and make a comparison with the time evolution of linear entropy and entanglement negativity. Particularly, we will compare the time scale for the saturation of complexity with the saturation time scale of linear entropy  and negativity. For this, it will be particularly useful to use what we name as the normalized complexity. Normalized complexity is the complexity divided by its saturation value.
\section{Open Quantum Systems}  \label{OQSDef}

In this section we will investigate the time evolution circuit complexity, linear entropy and negativity for two different open quantum systems. The first one is a two-oscillator model, which we will use as a toy model to investigate the saturation time scales for linear entropy, negativity and complexity. Then we will use the  Caldeira Leggett Model \cite{CALDEIRA1983374} to study the same.

\subsection{Two-oscillator Model} \label{2oscsec}
A very simple open quantum system to study the evolution of linear entropy, negativity and complexity is a model with two coupled oscillators \cite{open2}. We can think of one of the oscillator as the \textit{system} and the other oscillator as \textit{bath}. We now provide a brief review of the model following \cite{open2}.\par The Hamiltonian for such a model has the following form,
\begin{equation}
 H = \frac{1}{2} p_1^2 + \frac{\omega_0^2 \epsilon_1}{2} x_1^2
    + \frac{1}{2} p_2^2 + \frac{\omega_0^2 \epsilon_2}{2} x_2^2
    +  \omega_0^2 \lambda x_1 x_2  \ .
\label{Hsystem}
\end{equation}
In Eq.~\ref{Hsystem}, $x_i$ ($p_i$) is the position (momentum) operator at site-$i$,
with $[x_i, p_j] = i\, \delta_{i\,j}$; $\omega_0$ is a parameter with units of energy;
$\epsilon_1,\epsilon_2,\lambda $ are dimensionless real parameters.  
In what follows, we take $\lambda \geq 0$; 
we set $\omega_0=1$ i.e. $\omega_0$ sets the energy scale.
We are working in units where $\hbar = 1$.
The Hamiltonian in Eq.~\ref{Hsystem} can be easily diagonalized and take the following form. 
\begin{equation} \label{Hdiagonal}
 H = \frac{1}{2} P_s^2 + \frac{\Omega_s^2}{2} Q_s^2
    + \frac{1}{2} P_a^2 + \frac{\Omega_a^2}{2} Q_a^2  \end{equation}
where
$$\Omega_s^2 = \omega_0^2 \left[ (\epsilon_1 + \epsilon_2)/2 + E \right],\quad \Omega_a^2 = \omega_0^2 \left[ (\epsilon_1 + \epsilon_2)/2 - E \right]$$ are two normal mode frequencies, and $E = \sqrt{\epsilon_0^2 + \lambda^2}$, with $\epsilon_0 \equiv (\epsilon_1 - \epsilon_2)/2$.\par
%
We consider the following quench protocol,
\begin{subequations}
\begin{eqnarray}
 H & = & H_<  \ \ \ \ \ \ \ \ \ \ \  {\rm for} \ t < 0
  \\ 
 H & = & H_>  \ \ \ \ \ \ \ \ \ \ \  {\rm for} \ t > 0  \ ,
\nonumber 
\end{eqnarray}
where $H_<$ and $H_>$ have parameters 
\begin{eqnarray} \label{Hlessnew}
 & & H_<: \epsilon_1 = 1 \ , \ \epsilon_2 = 1 \ , \ \lambda = 0  \ ,
 \\
 & & H_>: \epsilon_1 \neq 1 \ , \ \epsilon_2 \neq 1 \ , \ \lambda > 0  \ . 
\nonumber
\end{eqnarray}
\end{subequations}
The system is initially in the ground state of $H_<$. Then we evolve the system with $H_>$ as $| \psi(t) \rangle = \exp(-i H_> t) | \psi_0 \rangle$. 
In our study, we will consider several different choices for the set of parameters that will lead to the following cases
\begin{itemize}
\item Both \textit{ordinary} oscillators : $\epsilon_1, \epsilon_2 > 0$. 
\item Bath (Oscillator-2) is \textit{inverted}, and system (Oscillator-1)
is \textit{normal} : $\epsilon_1 > 0;~ \epsilon_2 < 0$.
\item System (Oscillator-1)is \textit{inverted}, and Bath (Oscillator-2)
is \textit{normal} : $\epsilon_1 < 0;~ \epsilon_2 > 0$.
\end{itemize}
In the position representation, we can write the wave function as
\begin{subequations}
\begin{equation}
 | \psi(t) \rangle = \exp(-i H_> t) | \psi_0 \rangle 
 \longrightarrow
 \psi({\bf x},t) = \int d{\bf x}'~  K({\bf x},t | {\bf x}', t=0)~ \psi_0({\bf x}') \ ,
\end{equation}
where the initial wave function $\psi_0({\bf x})$ has the following form
\begin{equation}
 \psi_0({\bf x}) = \left( \frac{ {\rm det}( \hat{\Omega}_0 ) }{ \pi^2 } \right)^{1/4}
  \exp \left( - \frac{1}{2} {\bf x}^T \hat{\Omega}_0 {\bf x} \right)  \ ,
\end{equation}
and $K({\bf x}, {\bf x}' | t )$ is the propagator
\begin{equation}
 K({\bf x}, {\bf x}' | t ) = \langle {\bf x} | \exp(-i H_> t) | {\bf x}' \rangle
 \ .
\end{equation}
\end{subequations}
%
%
For the Hamiltonian given in Eq.~\ref{Hsystem}, the propagator can be easily computed as the Hamiltonian is of the quadratic form. Then carrying out the (Gaussian) integral, we obtain
\begin{equation}
 \psi({\bf x}, t) = \left ( \frac{ {\rm det}( {\rm Re}[\hat{\Omega}(t)] ) }{\pi^2} \right)^{1/4}
  \exp \left( - \frac{1}{2} {\bf x}^T \hat{\Omega}(t) {\bf x} \right)  \ ,
\label{wavefunctionmatrix}
\end{equation}
where
\begin{equation}  \label{newdef1}
 \hat{\Omega}(t) = \hat{G} ( \hat{\Omega}_0 - i \hat{F} )^{-1} \hat{G} - i \hat{F} 
 \,,
\end{equation}

\begin{subequations}
\begin{equation}
 \hat{F} = S  
  \left( \begin{array}{cc}
    f_s & 0 \\ 
    0 & f_a
    \end{array} \right) S^T
 \ \ , \ \ 
 \hat{G} = S  
  \left( \begin{array}{cc}
    g_s & 0 \\ 
    0 & g_a
    \end{array} \right) S^T
 \ \ , \ \ 
 S  =  \left( \begin{array}{cc}   
 u & -v \\ 
   v & u
   \end{array} \right)
\end{equation}
\end{subequations}
and 
\begin{align}
\begin{split}
&u = \sqrt{ \frac{1}{2} \left( 1 + \frac{\epsilon_0}{E} \right) }, \quad  v = \sqrt{ \frac{1}{2} \left( 1 - \frac{\epsilon_0}{E} \right) }, E = \sqrt{\epsilon_0^2 + \lambda^2}, \\&
f_{\alpha} = \omega_{\alpha} \cot (\omega_{\alpha} t), \ \ 
 g_{\alpha} = \omega_{\alpha}/\sin (\omega_{\alpha} t);\quad \text{where, } \alpha=a,s\,.
 \end{split}
 \end{align}

Explicitly, the wavefunction takes the following form, 
\begin{equation}
 \psi( x_{1}, x_{2} ) = \left ( \frac{ {\rm det}( {\rm Re}[\hat{\Omega}(t)] ) }{\pi^2} \right)^{1/4}
     \exp \left ( - \frac{1}{2} \left[ \Omega_{1}(t) x_{1}^2 + \Omega_{2}(t) x_{2}^2 
                    - \kappa(t) x_1 x_2 \right]  \right)  \ ,
\label{wavefunction}
\end{equation} 
where,  $\Omega_1(t), \Omega_2(t)$ and $\kappa(t)$ are the elements of the matrix defined in Eq.~\ref{newdef1} as defined below.
\begin{equation}
 \Omega_1(t) = \hat{\Omega}(t)_{11}  \ \ , \ \
 \Omega_2(t) = \hat{\Omega}(t)_{22}  \ \ , \ \ 
 \kappa(t) = - \hat{\Omega}(t)_{12}  \ .
\end{equation}\\
\textit{Linear entropy for Two-oscillator Model:} In what follows, as mentioned earlier we will consider \textit{oscillator-1} to be the \textit{system} and trace out the \textit{bath} i.e \textit{oscillator-2}.
The reduced density matrix $\hat \rho_1$ matrix for the system can be easily computed in the position space \footnote{We could have identified oscillator-2 as the system and oscillator-1 as the bath and traced out oscillator-1 to compute the reduced density matrix for oscillator-2. But the result presented in this paper will remain unchanged.}.
\begin{subequations}
\begin{equation}
 \rho_1(x_1,x_1') = \int dq_2~ \rho( x_1,q_2 \mid x_1',q_2) 
\label{rhoreduced}
\end{equation}
with
\begin{equation} 
 \rho(x_1,x_2 \mid x_1',x_2') = \psi(x_1,x_2) \psi^*(x_1',x_2')
\label{2particlerho}
\end{equation}
\end{subequations}
being the position-space density matrix of the full system i.e both the system (oscillator-1) and the bath (oscillator-2); the eigenvalues of 
$\hat{\rho}_1$ are obtained from the integral equation
\begin{equation}
 \int dx_1'~ \rho_1(x_1,x_1') \phi_n(x_1') = \rho_n \phi_n(x_1) \ .
\label{eigenproblem}
\end{equation}
After tracing out the oscillator-2 we get the following reduced density matrix
\begin{equation} \label{condition}
\rho_1 (x_1, x'_1) = \sqrt{\frac{\gamma_1 -\eta}{\pi} } \exp \left(-\frac{1}{2} (\gamma x_1^2+ \gamma^* x_1^{\p 2})+\eta\, x_1 x'_1 \right),
\end{equation}
where $\gamma(t) =\Omega_1(t) -\kappa(t)^2/2 \rm Re[\Omega_1(t)], \ \eta(t) =|\kappa(t)|^2/2 \rm Re[\Omega_2(t)].$
We will use this reduced density matrix to compute linear entropy as well as entanglement negativity and complexity subsequently.


To obtain the linear entropy first we note that the square of the density matrix operator in the position basis can be defined analogously,  
\begin{equation}
    \hat \rho_1^2 =\int dx dx' \rho_1^2(x_1, x'_1) |x_1\rangle \langle x'_1|,
\end{equation}
where $ \rho_1^2(x_1, x'_1)= \int dq \rho_{1}(x_1, q) \rho_1(q, x'_1)$. Then by computing the trace of the square of this reduced density matrix, we obtain the linear entropy as
\begin{equation} \label{compare1}
S_l=1- \rm Tr [\hat \rho^2_1] = 1- \int d x_1 \rho^2(x_1, x_1) = 1- \sqrt{\frac{\gamma_1 -\eta}{\gamma_1 +\eta}},
\end{equation}
where $\gamma_1$ is the real part of $\gamma$. \\

\noindent
\textit{Entanglement negativity for Two-oscillator Model:} Next we will compute the entanglement negativity for this two-oscillator model. First of all, the partial transpose of the density matrix (in the position basis) $\rho(x_1,x_2~ |~ x'_{1},x'_{2})$ of the full-system as shown in Eq.~\ref{2particlerho}, is defined in the following way,
\begin{align} \label{partialT}
\rho^{T_1}(x_1,x_2~ |~ x'_1,x'_2) &= \rho(x'_1,x_2~ |~ x_1,x'_2) \\
&= \psi(x'_1,x_2) ~\psi^*(x_1,x'_2)
\end{align}
As we like to focus on the dynamics of the \textit{system}, the partial transposition as defined in Eq.~\ref{partialT} has been done with respect to the oscillator-1.\par Since $\psi(x_1,x_2)$ is Gaussian, we can easily compute $\hat A=\sqrt{(\hat \rho^{ T_1})^\dagger \hat\rho^{T_1}}$ in the position basis.  Note that in the position basis $\hat A^2=(\hat \rho^{ T_1})^\dagger \hat\rho^{T_1}$,
\begin{equation}
\begin{split}
\hat A^2~ (x_1,x_2~ |~ x'_1,x'_2) & = \int dy_1 \, dy_2\, (\rho^{T_1})^\dagger (x_1,x_2~|~y_1,y_2) ~ \rho^{T_1} (y_1,y_2~|~x'_1,x'_2) \\
& = \tilde{{N}}\exp{\left(a_1 x_1^2 + b_1 x_1 x'_1 + c_1 x_1^{\p 2}+a_2 x_2^2 + b_2 x_2 x'_2 + c_2 x_2^{\p2}\right)},
\end{split}
\end{equation}
where $\tilde{{N}} = \left( \frac{ {\rm det}( {\rm Re}[\hat{\Omega}] ) }{\pi^2} \right)^{1/2}$ and \ $a_{1(2)} = \frac{(\kappa^*)^2}{4\, {\rm Re}{\,\Omega_{2(1)}}} - \frac{1}{2}\, \Omega_{1(2)}^*,\ b_{1(2)} = \frac{|k|^2}{2\, {\rm Re}{\,\Omega_{2(1)}}}, \ c_{1(2)} = a_{1(2)}^*. $\\

\noindent
Now, the square-root of this operator can be computed to be\footnote{For example, square-root of an operator $\mathcal{O} (x,x')$ can be evaluated by solving for $f(x,x')$, such that $\int dy\, f(x,y) f(y,x') = \mathcal{O}(x,x')$. If $\mathcal{O}(x,x')$ is a Gaussian function, $f(x,x')$ will also be a Gaussian function.},
\begin{align}
\begin{split}
&\hat A ~ \big(x_1,x_2~ |~ x'_1,x'_2\big) = {{N}}\exp{\left(A_1 x_1^2 + B_1 x_1 x'_1 + C_1 x_1^{\p 2}+A_2 x_2^2 + B_2 x_2 x'_{ 2} + C_2 x_2^{\p 2}\right)} ,
\end{split}
\end{align}
where the normalization factor
$ {{N}} = \left( \frac{{\rm det}( {\rm Re}[\hat{\Omega}] ) }{\pi^2} \right)^{1/2}$ and
\begin{equation}
A_{1 (2)} = \frac{1}{2}(2a_{1(2)}-b_{1(2)}), \ B_{1(2)} = \sqrt{2b_{1(2)}(b_{1(2)}-a_{1(2)}-c_{1(2)})}, \ C_{1(2)} =\frac{1}{2}(2c_{1(2)}-b_{1(2)}).
\end{equation}
Then we finally compute the trace of the above operator, as shown below:
\begin{align}
\|\rho^{T_1}\| &= \Tr{\sqrt{\left(\rho^{T_1}\right)^\dagger \rho^{T_1}}} \\
&=\int dx_1\,dx_2\,\sqrt{\left(\rho^{T_1}\right)^\dagger \rho^{T_1}} ~\big(x_1,x_2~ |~ x_1,x_2\big) \\
&= \left(\frac{{\rm det}( {\rm Re}[\hat{\Omega}])}{(A_1+B_1+C_1)(A_2+B_2+C_2)}\right)^{1/2}
\end{align}
Then by using  Eq.~\ref{negativity}, we get the entanglement negativity 
\begin{equation}
\mathcal{N}(\rho)=\frac{1}{2} \left[\left(\frac{{\rm det}( {\rm Re}[\hat{\Omega}])}{(A_1+B_1+C_1)(A_2+B_2+C_2)}\right)^{1/2} -1 \right] 
\end{equation}\\
\textit{COP for Two-oscillator Model:}
Now we will outline the computation for the COP. First, we will purify the reduced density matrix (by tracing out the oscillator-2) defined in Eq.~\ref{condition} in such a way that, 
\begin{equation} \label{cond}
\hat \rho_1= {\rm Tr}_{1'}|\psi_{11'} \rangle \langle \psi_{11'} |,
\end{equation}
where $1'$ corresponds to the auxiliary Hilbert-space such that $|\psi_{11'}\rangle$ is a pure state and $\hat \rho_1$ is the density matrix corresponding to the oscillator-1 as defined in Eq.~\ref{condition}. Then the complexity of purification ($\mathcal{C}_p$)  with respect to the ground state of $H_{<}$ in Eq.~\ref{Hlessnew} (with $\epsilon_1=\epsilon_2=\omega_0=1$ and $\lambda=0$) as the reference state is defined as,
\begin{equation} \label{min}
\mathcal{C}_p= \min_{1'}\mathcal{C}(|\psi_{11'}\rangle).
\end{equation}
Here the minimization is over all possible purification and $\mathcal{C}(|\psi_{11'}\rangle)$ corresponds to the complexity of the state $|\psi_{11'}\rangle $.  Note that, here we choose the  minimal Gaussian purification such that the size of the original (sub) system is same as the auxiliary system as in \cite{Caceres:2019pgf,Bhattacharyya:2018sbw, Bhattacharyya:2019tsi}. In our case the subsystem is just one oscillator.

Next we parametrize our purified state in the following way, 
\begin{equation} \label{pur}
\psi(x_1,x_2)= {N}\exp \left (-\frac{1}{2} \Big[\alpha x_1^2+\gamma x_2^2-2\, \tau x_1 x_2\Big]\right ),
\end{equation}
where $x_2$ belongs to the auxiliary Hilbert-space. The parameters $\alpha,\gamma $ and $\tau$ are in general complex and yet to be determined.  ${N}$ is the normalization of the wavefunction. Then we have
\begin{equation}
\rho(x_1, x_2 \, | \, x_1', x_2')= \psi(x_1,  x_2) \psi^{*} (x_1', x_2')
\end{equation}
The corresponding reduced density matrix after tracing out the auxiliary Hilbert space is
\begin{align}
\begin{split}
&{\rm Tr}_{2}\,\rho(x_1, x_2 \, | \, x_1', x_2')= \int_{-\infty}^{\infty} dx_2 \, \psi(x_1,  x_2) \psi^{*} (x_1', x_2)\\&
= \displaystyle{{N}^2\exp\left (-\frac{1}{2}\left[\left(\alpha-\frac{\tau^2}{2\, {\rm Re}(\gamma)}\right)x_1^2+\left(\alpha^*-\frac{(\tau^*)^2}{2\, {\rm Re}(\gamma)}\right) x_1'^2\right]+\frac{|\tau|^2}{2\,{\rm Re}(\gamma)} x_1 x_1'\right)}
\end{split}
\end{align}
Using the condition mentioned in Eq. \ref{cond} and using Eq. \ref{condition} we get the following,  
\begin{equation} \label{apr}
\alpha= \Omega_1(t), \tau= \kappa (t), {\rm Re }(\gamma)= {\rm Re }(\Omega_2(t)).
\end{equation}

We have determined all the parameters inside Eq. \ref{pur} in terms of the given parameters as in Eq. \ref{apr}, except for ${\rm Im }(\gamma)$. Hence the minimization in (\ref{min}) can be carried over  ${\rm Im }(\gamma)$ and the minimum value will correspond to the complexity of purification. We compute the complexity corresponding to Eq. \ref{pur} as follows:
\begin{equation}
\mathcal{C}(|\psi_{11'}\rangle)=\frac{1}{2} \sqrt{ \sum_{i=1}^{2} \left[
     \ln \left (\frac{| \omega_{i}|}{\omega_0} \right)^2 
  + \tan^{-1} \left (-\frac{ \text{Im} ( \omega_{i})}{\text{Re} ( \omega_{i})} \right)^2 
  \right] }  \ ,
\end{equation}
where, 
\begin{align}
\begin{split}
\omega_1=\frac{1}{2}(\alpha+\gamma+\sqrt{(\alpha-\gamma)^2+ 4\tau^2}), \ \omega_2=\frac{1}{2}(\alpha+\gamma-\sqrt{(\alpha-\gamma)^2+ 4\tau^2}),
\ \omega_0=1.
\end{split}
\end{align}
Finally, the complexity of purification is
\begin{equation} \label{compare2}
\mathcal{C}_p=\min_{{\rm Im }(\gamma)}\Bigg(\frac{1}{2} \sqrt{ \sum_{i=1}^{2} \left[
     \ln \left (\frac{| \omega_{i}|}{\omega_0} \right)^2 
  + \tan^{-1} \left (-\frac{ \text{Im} ( \omega_{i})}{\text{Re} ( \omega_{i})} \right)^2 
  \right] } \Bigg).
\end{equation}\\
\textit{Complexity from Operator-State Mapping for Two-oscillator Model:}
Finally we will compute the complexity by using the operator state mapping. 
Motivated by the thermofield-double state \cite{Hosur:2015ylk}, we work with $\hat{\rho}^{1/2}_1$ where again the density matrix $\rho_1$ for the oscillator-1 is defined in Eq.~\ref{condition}; working in the position representation, one has an effective wave function (in this doubled Hilbert space)
\begin{equation}
 \psi(x,x') =    \frac{1}{ \sqrt{ {\rm Tr}[ ( \hat{\rho}^{1/2}_1 )^{\dagger} 
      \hat{\rho}^{1/2}_1 ] } }  \rho^{1/2}_1(x',x)  \ . 
\label{effectivepsi}
\end{equation}
We will use this as the target state and compute complexity. We can write the wavefunction explicitly as follows \cite{open2}:
\begin{eqnarray} \label{effectivepsi1}
 \psi(x_1',x_1) & = & {N}
   \exp \left (- \frac{1}{2} (\gamma + \eta) x_1^2 - \frac{1}{2} (\gamma^* + \eta) x_1'^2 
          + \sqrt{2 \eta (\gamma_1 + \eta) } x_1 x'_1 \right ) \ .
\end{eqnarray}
Here ${N}$ is a normalization constant and $\gamma_1$ is the real part of $\gamma$. 
The effective wave function can be written in the form
\begin{equation}
 \psi(x_1',x_1) = {N}  \exp \left ( - \frac{1}{2} \left( \beta x_1^2 + \beta^* x_1'^2 \right)
      + \zeta  x_1 x_1' \right)  \ .
\end{equation}
To proceed, we need to diagonalize the argument of the exponential; we obtain the effective wave function
\begin{subequations}
\begin{equation}
 \psi(x',x) = { N} 
  \exp \left (- \frac{1}{2} (\beta_1 + E) X_1^2 - \frac{1}{2} (\beta_1 - E) X_2^2 \right )
  \label{state2}
\end{equation}
where 
\begin{equation} 
 \left( \begin{array}{c}
    X_1 \\ X_2
    \end{array} \right) 
 = \left( \begin{array}{cc}
    u & -v \\ 
    v & u
    \end{array} \right) 
 \left( \begin{array}{c}
    x_1 \\ x'_1 
    \end{array} \right) 
    \label{waveeffective}
    \end{equation}
 and
 \begin{equation}
 E^2 = \zeta^2 - \beta_2^2 \ \ , \ \
 u = \sqrt{ \frac{1}{2} \left( 1 + i \ \frac{\beta_2}{E} \right) }
 \ , \ 
 v = - \sqrt{ \frac{1}{2} \left( 1 - i \ \frac{\beta_2}{E} \right) }  \ ,
\end{equation}
\end{subequations}
with $\beta_1 = {\rm Re}[\beta]$, $\beta_2 = {\rm Im}[\beta]$.
In what follows, we will use this effective wavefunction (\ref{waveeffective}) as the target state and compute the complexity [see Appendix~\ref{A}] with respect to the ground state wavefunction of $H_<$ defined in Eq.~\ref{Hlessnew}.

\subsubsection{Time Evolution of Linear Entropy, Negativity and Complexity}

\begin{figure}[t!]
\begin{center}
\scalebox{1.2}{\includegraphics{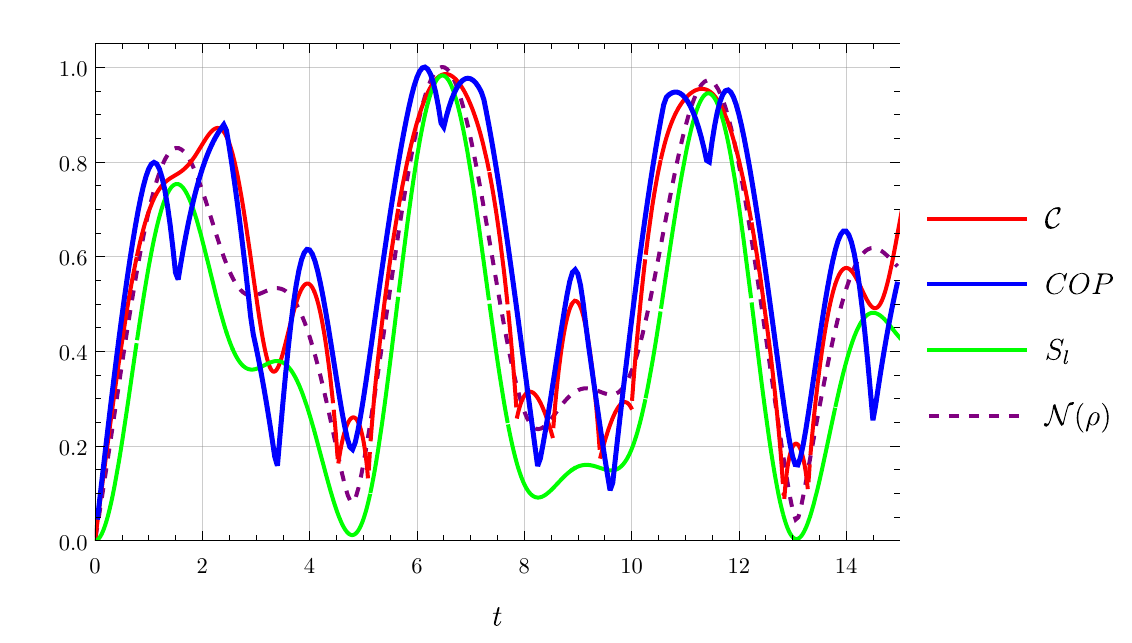} }\\
\vspace{-0.1in}
\caption{Complexity from operator-state mapping ($\mathcal{C}$), COP, Linear Entropy  and Entanglement Negativity vs time for small $\lambda$ (i.e $\lambda < 1$) when both the system and bath are regular oscillators ($\epsilon_1 = \epsilon_2 = 1$ and $ 
\lambda = 0.5$).
 } \label{CvsPm1}
\end{center}
\end{figure}

We study and compare the time evolution of the subsystem complexity (both COP and complexity from operator-state mapping), linear entropy and entanglement negativity for different choices of system parameters. 
Below we list the main features and conclusion drawn from them. 
\begin{itemize}
    \item When both the system and bath are ordinary oscillators, all four quantities show oscillatory behavior and do not saturate as evident from Fig.~\ref{CvsPm1}. Moreover, the peaks appear around the same time for all the quantities. 


\item Entanglement negativity $\mathcal{N}(\rho)$ does not saturate for any values of coupling $\lambda$. Apart from the case shown in Fig.~\ref{CvsPm1}, when both the system and the bath is regular and the coupling is small ($\lambda < 1 $), it always  grows exponentially with time as shown in Fig.~\ref{EnNegAll}. This can be contrasted with the behaviour of the linear entropy and complexity. When both the system and the bath are regular, then irrespective of the magnitude of $\lambda$ both the complexity and the linear entropy shows an oscillatory behaviour as demonstrated in the Fig.~\Ref{CvsPm1} and Fig.~\ref{CvsPm2}. 
\begin{figure}[htb!]
\begin{center}
\scalebox{1.15}{\includegraphics{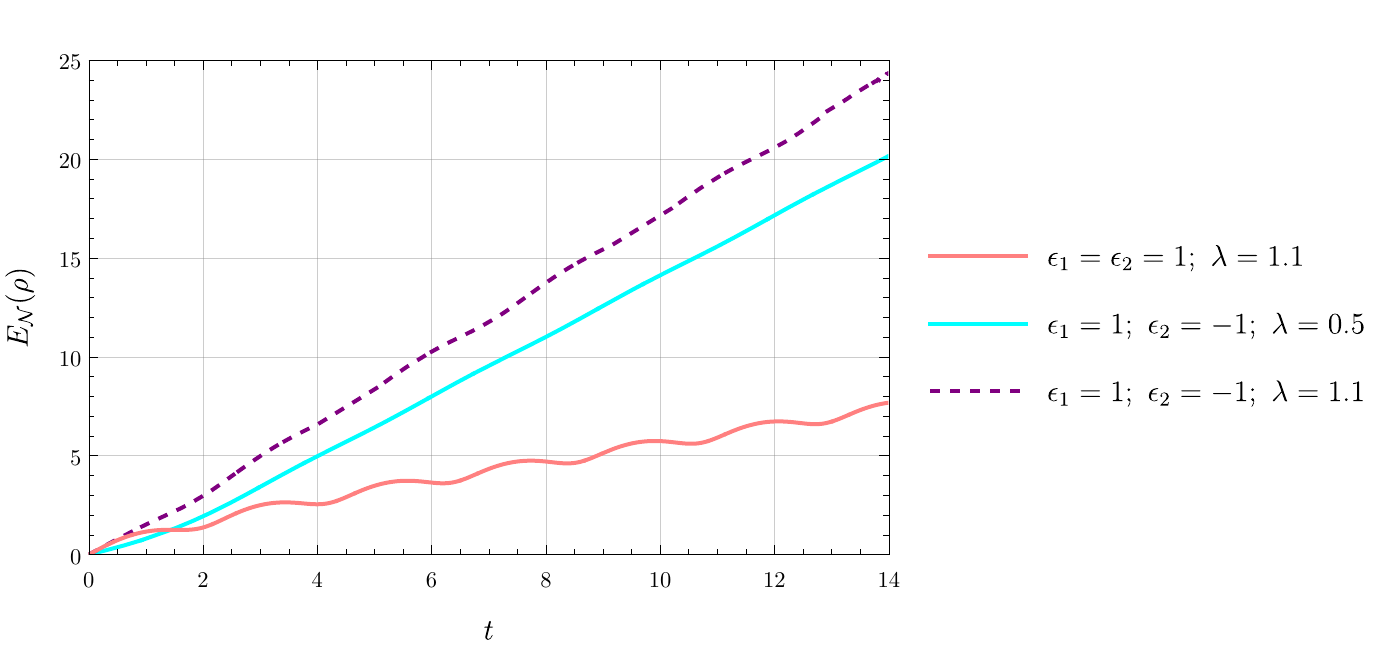} }
\caption{Logarithmic Negativity vs time for different parameter choices in two-oscillator model (Note that, the entanglement negativity remains the same even if we exchange the values $\epsilon_1 \leftrightarrow \epsilon_2$.) } \label{EnNegAll}
\end{center}
\end{figure}
    \item When either the bath or the system is inverted, both the  complexity and the linear entropy reach to a saturation as shown in the Fig.~\ref{CvsPm3} and Fig.~\ref{CvsPm5}. Interestingly both complexities (COP and complexity from state-operator mapping) reach to their saturation values \footnote{Both complexities have oscillations around the saturation value, which is an expected feature for harmonic oscillator models.} faster than the linear entropy
    as evident from the Fig.~\ref{CvsPm3} and Fig.~\ref{CvsPm5}. \textit{Moreover, we can also conclude that when the system becomes \textit{mixed}, the complexity is expected to be saturated.}  We have extensively scanned the parameter space for this two-oscillator model. 
   

\begin{figure}[!]
\begin{center}
\scalebox{1.1}{\includegraphics{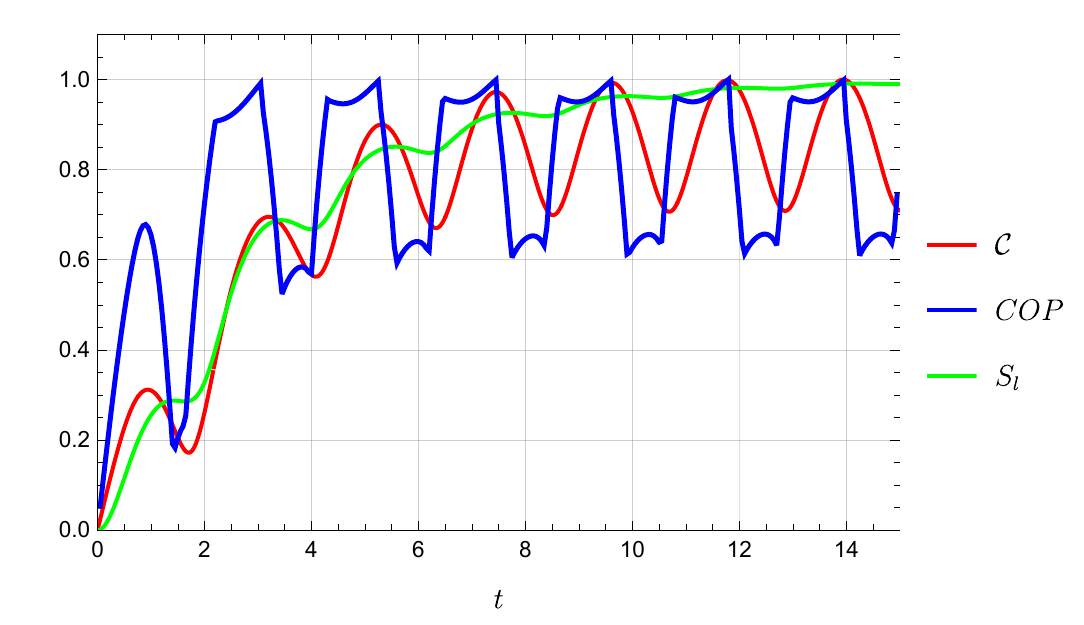} }
 \\
\vspace{-0.1in}
\caption{Complexity from operator-state mapping ($\mathcal{C}$), COP and Linear Entropy vs time for large $\lambda$ (i.e $\lambda > 1$) when both the system and bath are regular oscillators ($\epsilon_1 = \epsilon_2 = 1$ and $ \lambda = 1.1$).}
 \label{CvsPm2}
\end{center}
\end{figure}



\item 
Last but not the least, the exponential growth of entanglement negativity (or the linear growth of the logarithmic negativity as shown in Fig.~\ref{EnNegAll}) \footnote{It can be easily checked that that, the entanglement entropy also grows with time when either of the oscillator becomes inverted.} can be contrasted  with the behaviour of both the complexities (from operator-state mapping and COP). The complexity saturates when either of the oscillator is inverted and the system gets maximally mixed. Note that the two-oscillator model is only a toy model. In the next section, we will see that for a more realistic model (Caldeira Leggett model) negativity will also saturate when the system becomes mixed.

\begin{figure}[htbt!]
\begin{center}
\includegraphics[width=1.0\linewidth]{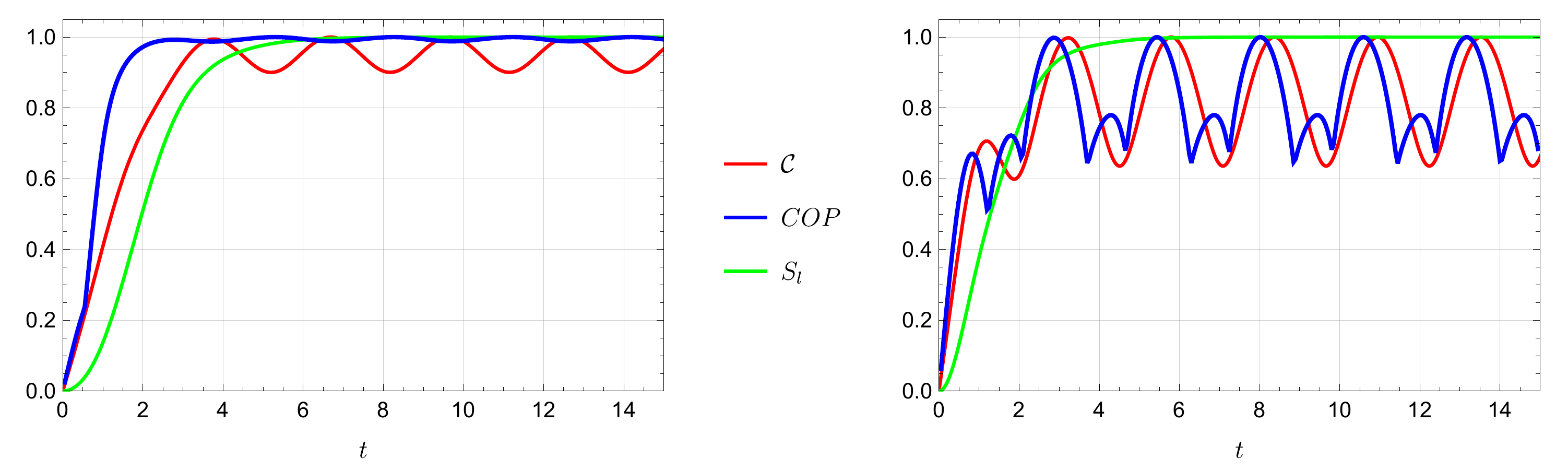}
\caption{Complexity from operator-state mapping ($\mathcal{C}$), COP and Linear Entropy  vs time for small $\lambda$ when
bath is inverted and system is normal 
($\epsilon_1 = 1,\,  \epsilon_2 = -1)$ and $ 
\lambda = 0.5$ (Left) $\lambda = 1.2$ (Right). } \label{CvsPm3}
\end{center}
\end{figure}

\begin{figure}[htbt!]
\begin{center}
\includegraphics[width=1.0\linewidth]{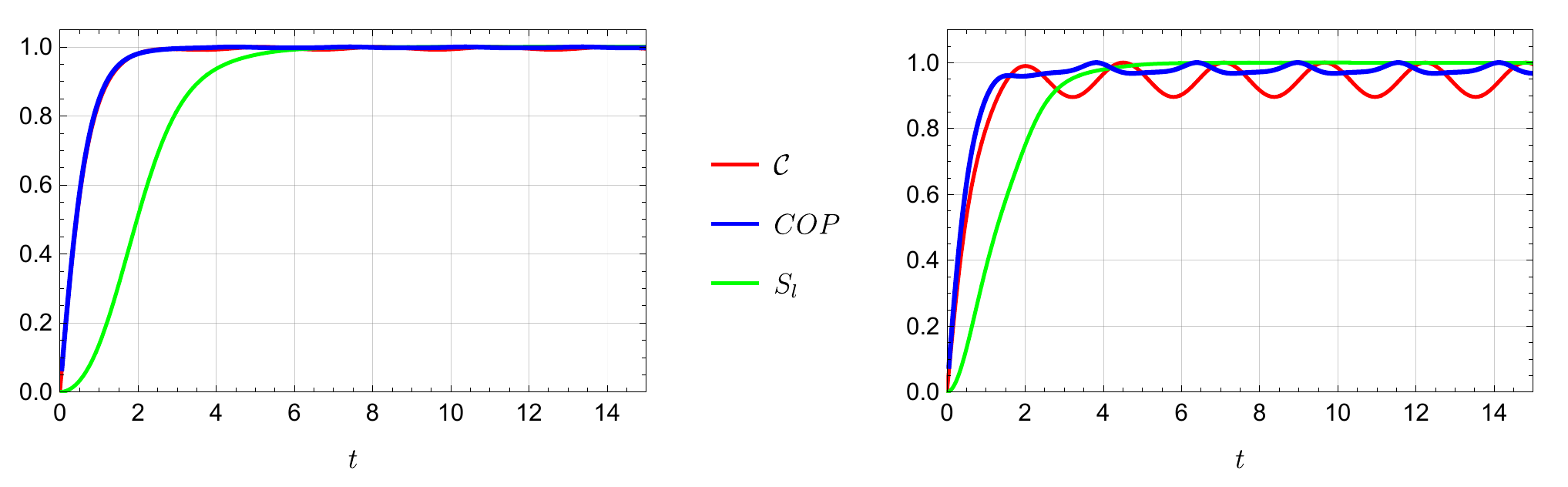}
\caption{Complexity from operator-state mapping ($\mathcal{C}$), COP and Linear Entropy vs time for small $\lambda$ when
system is inverted and bath is normal 
($\epsilon_1 = -1,\,  \epsilon_2 = 1)$ and $ 
\lambda = 0.5$ (Left) $\lambda= 1.1$ (Right). } \label{CvsPm5}
\end{center}
\end{figure}

\end{itemize}





\subsection{Caldeira Leggett Model}
We now turn our attention to yet another simple model of open quantum system, namely the \textit{Caldeira Leggett} model. The Caldeira Leggett model \cite{CALDEIRA1983374,PhysRevLett.46.211} provides one of the first microscopic descriptions of Quantum Brownian motion. It's a simple model where the dissipative dynamics of a harmonic oscillator interacting with a bosonic bath is captured by a position-position coupling. Below we review some details of our setup following \cite{PhysRevD.105.046011}. Interested readers are referred to \cite{PhysRevD.105.046011} for more details.\par

\noindent
\textit{Solving the Model:}
In principle, there are two ways to model such a heat bath. One way is is to consider a reservoir composed of a large ensemble of non-interacting quantum systems such as harmonic oscillators and the second way is to use a free field.  In this paper we will consider the second approach as in \cite {zurek0,PhysRevD.105.046011}.
Explicitly, the system we consider is a harmonic oscillator that is coupled to a one-dimensional free bosonic field. The Hamiltonian is given by
\begin{equation}
 H = \int_0^L \hspace{-0.1in} dx~ \left\{ 
    \frac{1}{2} \left[ \Pi^2 + (\partial_x \phi)^2 \right]
 + \delta(x) \left[ \frac{1}{2} P^2 + \frac{\omega^2_0}{2} Q^2 
      + \lambda Q \partial_x \phi  \right]  \right\} \ ,
\label{Hcanonical}
\end{equation}
where we are ultimately interested in the $L \rightarrow \infty$ limit.
In Eq.~\ref{Hcanonical}, 
$Q$ and $P$ are canonically conjugate variables describing the system, $[Q,P] = i$;
$\phi$ and $\Pi$ are canonically conjugate fields of the bath, 
$[ \phi(x), \Pi(x') ] = i \delta(x - x')$.\footnote{We work in units where $\hbar=1$.  
We have also set the string's speed of sound to unity: $v=1$.}
Furthermore, the field $\phi$ satisfies Dirichlet boundary conditions at $x=0$ and $x=L$:
\begin{align}
\phi(x=0)=0;\hspace{1cm}  \phi(x=L)=0. \label{bcphi}
\end{align}
For further simplification, we define the \textit{decay rate}, $\Gamma \equiv \lambda^2/2$. 
We consider the following quench in the above model (Eq.~\ref{Hcanonical}) --- 
\begin{equation} \label{quenchnewCL}
 H = \left\{ \begin{array}{c c c}
    H_<  & {\rm for}  &  t < 0  \\
    H_>  & {\rm for}  &  t > 0
    \end{array} \right.  \ ,
\end{equation}
In what follows, we start the system in the ground state of $H_<.$ We take $H_<$ to be the Hamiltonian of the system and bath decoupled. 
\begin{equation}
 H_< = \int_0^L \hspace{-0.1in} dx~
    \frac{1}{2} \left[ \Pi^2 + (\partial_x \phi)^2 \right]
 + \frac{1}{2} P^2 + \frac{\Omega_<^2}{2} Q^2 
\label{Hinit}
\end{equation}
By performing the following mode expansion \cite{PhysRevD.105.046011},
\begin{eqnarray}
 \phi(x) &=& \sqrt{\frac{2}{L}} \sum_{n=1}^{\infty} \sin \left( \frac{\pi n}{L} x \right)\,
 \frac{1}{\sqrt{2\Omega_n} } ( a^{\phantom \dagger}_n + a^{\dagger}_n )\,,
 \cr
 \Pi(x) &=& -i \sqrt{\frac{2}{L}} \sum_{n=1}^{\infty} \sin \left( \frac{\pi n}{L} x \right)\,
 \sqrt{\frac{\Omega_n}{2} } ( a^{\phantom \dagger}_n - a^{\dagger}_n ) \cr
 Q &=& \frac{1}{\sqrt{2\Omega_<} } ( a_0 + a_0^{\dagger} ) \cr
 P &=& -i \sqrt{ \frac{\Omega_<}{2} } ( a_0 - a_0^{\dagger} ) \ ,
\label{modedirichlet}
\end{eqnarray}
we get
\begin{equation}
 H_< = \sum_{n=1}^{\infty} \frac{\pi\, n}{L} \left( a^{\dagger}_n a^{\phantom \dagger}_n + 1/2 \right)
 + \Omega_< \left( a_0^{\dagger} a_0^{\phantom \dagger} + 1/2 \right)\,.
\nonumber
\end{equation}
Our initial state $| \psi_0 \rangle$ is annihilated by the $\{ a_i \}$: 
\begin{equation} \label{initialwavefunctioncl}
 a_i \mid \psi_0 \rangle = 0 \ \  \forall \  i\,.
\end{equation}

Then our final state is obtained by evolving $| \psi_0 \rangle$ with $H_>$
and $H_>$ is given by Eq.~\ref{Hcanonical}.
\begin{equation} \label{timeevolvedwave}
    |\psi(t)\rangle =e^{-i\,H_{>}\,t} |\psi_0\rangle.
    \end{equation}

\par  We will consider two different sets of parameters:
\begin{itemize}
\item \textit{Underdamped} oscillator : $\Omega^2 > 0$, i.e, $\omega_0^2 > \Gamma^2$.
\item \textit{Overdamped} oscillator : $\Omega^2 < 0$,  i.e, $\omega_0^2 < \Gamma^2$.
\end{itemize}
%
\noindent
To proceed further we first introduce a dual field $\theta$  such that 
\begin{equation} \Pi = - \partial_x \theta \,,
\end{equation}
with $[ \phi(x), \theta(x') ] = (i/2) {\rm sgn}(x' - x)$. In terms of these fields now the bath Hamiltonian takes the form \cite{PhysRevD.105.046011},
\begin{equation}
 H_B = \int_0^L \hspace{-0.1in} dx~ 
    \frac{1}{2} \left[ (\partial_x \theta)^2 + (\partial_x \phi)^2 \right] \ .
\nonumber
\end{equation}
Now we decompose $\phi$ and $\theta$ into right and left movers
\begin{subequations}
\begin{equation}
 \phi(x) = \phi_R(x) + \phi_L(x)
 \ \ , \ \ 
 \theta(x) = \phi_R(x) - \phi_L(x)  \ ,
\end{equation}
where $\phi_R$ and $\phi_L$ satisfy the commutation relations
\begin{equation}
 [ \phi_R(x), \phi_R(x') ] = \frac{i}{4} {\rm sgn}(x-x')
 \ \ , \ \ 
 [ \phi_L(x), \phi_L(x') ] = -\frac{i}{4} {\rm sgn}(x-x')
 \ \ , \ \ 
 [ \phi_R(x), \phi_L(x') ] = 0 \ .
\end{equation} 
\end{subequations}
\par 
In the Heisenberg picture, 
$\phi_R$ is a function of $(x-t)$, and $\phi_L$ is a function of $(x+t)$:
$\phi_R(x,t) = \phi_R(x-t)$ and $\phi_L(x,t) = \phi_L(x+t)$.
Using the Dirichlet boundary condition $\phi(x=0,t)=0$, we have $ \phi_L(x=0,t) = - \phi_R(x=0,t)\,.$
This implies
\begin{subequations}
\begin{equation}
 \phi_L(x+t) = - \phi_R(-x-t) \ ,
\end{equation}
namely that  $\phi_L$ can be regarded as the continuation of $\phi_R$ to $x < 0$.
Then using the boundary condition $\phi(x=L,t)=0$, one obtains
\begin{equation}
 \phi_R(-L,t) = \phi_R(L,t) \ ,
\end{equation}
\end{subequations}
so that we can work with right-movers on the interval $-L < x < L$ satisfying periodic boundary conditions. For more details please refer to the  Appendix~\ref{A}.
Writing the full Hamiltonian defined in Eq.~\ref{Hcanonical} solely in terms of right movers, we obtain \cite{PhysRevD.105.046011},
\begin{equation}
H = \int_{-L}^{L} \hspace{-0.1in} dx~ \left\{ (\partial_x \phi_R)^2
   + \delta(x) \left[ \frac{1}{2} P^2 + \frac{\omega^2_0}{2} Q^2 
      + 2 \lambda Q \partial_x \phi_R  \right]  \right\} \,.
\label{Hchiral}
\end{equation}
From Eq.~\ref{Hchiral}, we obtain the (Heisenberg) equations of motion for the operators
\begin{equation}
  \partial_t \phi_R = - \partial_x \phi_R - \delta(x) \lambda Q
\ \ , \ \ 
 \frac{d^2 Q}{dt^2} + \omega^2_0 Q = -2 \lambda \partial_x \phi_R(0)
 \ .
\label{heisenberg}
\end{equation}
The homogeneous part of the first equation in Eq.~\ref{heisenberg} tell us that, the  $\phi_R$ is a function of $(x-t)$; using that as a ``source" term acting only at $x=0$, we integrate about a small
region about $x=0$: $-\epsilon<x<+\epsilon$.
Eq.~\ref{heisenberg} becomes
\begin{eqnarray}
0 &=& \phi^+_R(x=+\epsilon) - \phi^-_R(x=-\epsilon)\Big|_{\epsilon\rightarrow 0} + \lambda\, Q \nonumber \\
 \frac{d^2 Q}{dt^2} + \omega^2_0 Q &= &
  - \lambda \left[ \partial_x \phi^+_R(x=0) + \partial_x \phi^-_R(x=0) \right]
 \ ,
\label{scattering}
\end{eqnarray} 
where we have introduced the notation
$\phi^+_R(x) \equiv \phi_R(x>0)$ and $\phi^-_R(x) \equiv \phi_R(x<0)$.


We will apply the scattering formalism \cite{buttiker} to solve Eq.~\ref{scattering}. In this formalism, we consider particles incoming from $x < 0$, then they are scattered by the system at $x=0$, and finally are outgoing for $x > 0$.
From the practical point of view, we solve for $Q$ and $\phi^+_R$ in terms of $\phi^-_R$. Finally we get
\begin{equation}
    \frac{d^2 Q}{dt^2} + \omega_0^2 Q  +2 \Gamma\, \frac{d Q}
    {dt}= 2 \sqrt{2\Gamma}\,\frac{d\phi_R^{-}}{dt},
\end{equation}
where  $\Gamma = \lambda^2/2$. 
The solution for $Q(t)$ can be written as
\begin{equation}
    Q(t)= Q_H(t)+Q_P(t),
\end{equation}
where, $Q_H(t)$ and $Q_P(t)$ are the homogeneous and particular solutions respectively. The homogeneous solution turns out to be the following:
\begin{eqnarray} \nonumber
 Q_H(t) & = & e^{-\Gamma t} \left\{ [ \cos \Omega t + \left(\frac{\Gamma}{\Omega} \right)\sin \Omega t ] Q(t=0)
        +(1/\Omega) \sin \Omega t~ P(t=0)  \right\} 
  \label{solutionQ} \\
        & - & e^{-\Gamma t} \left\{ [ \cos \Omega t + \left(\frac{\Gamma}{\Omega} \right) \sin \Omega t ] Q_P(t=0)
        + (1/\Omega) \sin \Omega t~ \dot{Q}_P (t=0) \right\}  \ ,
 \end{eqnarray} 
 where $\Omega= \sqrt{\omega_0^2-\Gamma^2}$. To find the particular solution we introduce the Fourier transforms\footnote{At this point, we have taken $L \rightarrow \infty$.}
\begin{equation}
 \phi_R(x,t) = \int \hspace{-0.069in} \frac{d\omega}{\sqrt{2\pi} } e^{-i \omega (t-x) } \phi_R(\omega)
 \ \ , \ \ 
 Q(t) = \int \hspace{-0.069in} \frac{d\omega}{\sqrt{2\pi} } e^{-i \omega t} Q(\omega)  \ ,
\end{equation}
and obtain
\begin{equation}
\begin{aligned}
 Q_P(\omega) & =  \left( \frac{ i 2 \sqrt{2\Gamma} \omega}
     { \omega^2 + i 2\Gamma \omega - \omega_0^2} \right) \phi_R^{-}(\omega)  \ ,  
  \\  
 \phi_R^{+}(\omega) &= - \sqrt{2\Gamma} \ Q_H(\omega) 
      + \left( \frac{ \omega^2 - i 2 \Gamma \omega - \omega_0^2}
          { \omega^2 + i 2\Gamma \omega - \omega_0^2} \right) \phi_R^{-}(\omega)  \ .
\label{solutionR}
\end{aligned}
\end{equation}

\noindent
To execute calculations for the bath, one must understand $\phi_R^-(\omega)$ as
\begin{equation}
 \phi_R^-(\omega) = \left\{ \begin{array}{c c c}
   \frac{a^{\phantom \dagger}_{\omega}}{\sqrt{2\omega}} & {\rm for} & \omega > 0  \\
   \frac{ a^{\dagger}_{\omega}}{\sqrt{2|\omega|}} & {\rm for} & \omega < 0  \\ 
    \end{array} \right.  \ ,
\nonumber
\end{equation}
namely, that $\phi_R^{-}(\omega)$ is an annihilation operator for $\omega > 0$ 
and a creation operator for $\omega < 0$ (with a prefactor $1/\sqrt{2|\omega|}$). Finally, the bosonic bath fields $\phi(x,t)$ and $\Pi(x,t)$ can be expressed as:
\begin{align} \label{solutionnew}
\phi(x,t) &= \phi_0(x,t) - {\rm sgn}{(x)}\,\lambda Q(t-|x|) \\
\Pi(x,t) &=  \Pi_0(x,t) - {\rm sgn}{(x)}\,\lambda \dot{Q}(t-|x|) 
\end{align}
where, $\phi_0(x,t)$ and $\Pi_0(x,t)$ are the time evolution of the bosonic fields in the \textit{free} theory, i.e,
\begin{equation}
\begin{aligned}
\phi_0(x,t) &\equiv -i \,\int_0^{\infty} \frac{d\omega}{\sqrt{\pi \omega}}\, \sin{(\omega x)}\,\left(a_{\omega}^\dagger e^{i\omega t} - a_{\omega} e^{-i\omega t}\right)\,,  \\
\Pi_0(x,t) &\equiv  \int_0^{\infty} d\omega\, \sqrt{\frac{\omega}{\pi}}\, \sin{(\omega x)}\,\left(a_{\omega}^\dagger e^{i\omega t} + a_{\omega} e^{-i\omega t}\right)  \,.
\end{aligned}\label{bosdef}
\end{equation}
After knowing the solutions mentioned in Eq.~\ref{solutionQ}, \ref{solutionR} and \ref{solutionnew}, we are now in a position to compute the time-evolved wavefunction mentioned in Eq.~\ref{timeevolvedwave}. \\\\
\textit{Reduced Density Matrix for the Oscillator (system):} In this paper we are interested in the reduced density matrix of the system by tracing out the bath degrees of freedom. To obtain that, we divide the full system into two subsystems-oscillator and the string 
and then traced out the string. 
\begin{equation}
 \hat{\rho_S} = {\rm Tr}_B[ \hat{\rho} ] \ , 
\end{equation}
where $\hat{\rho}$ is the density matrix of the entire system. The subscript $S$ and $B$ implies system and bath respectively. 
%
The system starts from the ground state of the uncoupled Hamiltonian defined in Eq.~\ref{Hinit}. Then we time evolve it by $H_{>}$ defined in Eq.~\ref{quenchnewCL}. As the initial density matrix is Gaussian and the Hamiltonian is quadratic, the density matrix remains Gaussian for all times. In the position representation, it has the following structure
\begin{equation}
 \rho_S(x,x') = \sqrt{\frac{\gamma_1 - \eta}{\pi} }
   \exp \left \{ - \frac{\gamma}{2} x^2 -\frac{\gamma^*}{2} x'^2 + \eta\,l x x' \right \} \ ,
\label{rhoSstructure}
\end{equation}
where $\gamma \ (= \gamma_1 + i \gamma_2) \in {\bf C}$ and $\eta \in {\bf R}$.

The parameters $\{\gamma, \eta \}$ can be obtained from the system's correlation functions. 
As the density matrix is Gaussian, it is fully characterized by its first and second moments. It can be easily checked that, the initial wave function mentioned in Eq.~\ref{initialwavefunctioncl}, the folloiwng expectation values vanish. $$\langle Q \rangle =0\quad \textrm{and}\quad \langle P \rangle = 0.$$ Then due to the structure of the Hamiltonian, this is preserved for all times. The  parameters in the density matrix defined in  Eq.~\ref{rhoSstructure} are given in terms of two point correlators of canonical variables $Q$ and $P$ describing the single oscillator \cite{PhysRevD.105.046011},

\begin{equation}
 \gamma_1 = \sigma_P - \frac{1}{\sigma_Q} \left( \sigma_{QP}^2 - 1/4 \right)
 \ \ , \ \ 
 \gamma_2 = - \frac{\sigma_{QP} }{\sigma_Q}
 \ \ , \ \ 
 \eta = \sigma_P - \frac{1}{\sigma_Q} \left( \sigma_{QP}^2 + 1/4 \right)  \ ,
\end{equation} 
where
\label{correlator}
\begin{equation} 
 \sigma_{Q} = \langle Q(t) Q(t) \rangle  \ \ , \ \ 
 \sigma_{P} = \langle P(t) P(t) \rangle  \ \ , \ \  
 \sigma_{QP} = \frac{1}{2} \langle Q(t) P(t) + P(t) Q(t) \rangle  \ \ ;
\end{equation}
Interested readers are referred to Appendix~\ref{B} and \ref{C} for the details of the computation of these correlators. \\

\noindent
\textit{Linear Entropy, Negativity and Complexity for Caldeira Leggett Model:}
Once we have the $\gamma$ and $\eta$ as defined in Eq.~\ref{rhoSstructure}, we follow the same procedure outlined for the two oscillator case in Sec.~\ref{2oscsec} to compute linear entropy, negativity and complexities. Linear entropy is $S_l$ is given for this case by the Eq.~\ref{compare1} 
\par
Next we compute the entanglement negativity following the procedure as discussed in Section \ref{EntropyDef} for two-mode Gaussian state. 
First we need to compute first the density matrix for the full system (i.e for oscillator and the bath together). Since this is also of the Gaussian form it is completely characterised by its first and second moments as mentioned earlier. All the first moment vanishes because of our choice of the initial state as mentioned earlier. Then we compute the covariance matrix as defined in Eq.~\ref{cvdef}. Note that the term $\delta(x)\, \lambda \, Q\,\partial_x \phi $ in the Hamiltonian (Eq.~\ref{Hcanonical}) denotes the interaction between the bosonic bath $\phi(x)$ and the harmonic oscillator $Q$ to be only at $x = 0$. Therefore, we consider the following canonical phase-space variables describing the system and bath, $$X_i = \{Q,\,P,\,{\phi(x=0)},\,{\Pi(x=0)}\}.$$
We have already computed the correlators corresponding to system variables $ Q(t), P(t)$. 
To compute the negativity, in addition to the system correlators, we now need to compute the two point correlators of system-bath and bath variables. However, the boundary conditions on $\phi(x)$ defined in Eq.~\ref{bcphi} sets ${\phi(0)} = {\Pi(0)} = 0$, and the corresponding correlators become zero. As a result, the negativity becomes undefined. Therefore we introduce a small length parameter $\delta \ll L$, and consider the bosonic fields at $x= \delta$. Then we can compute the correlators between the phase-space variables $\{Q,\,P,\,{\phi(x=\delta)},\,{\Pi(x=\delta)}\}$ (see Appendix \ref{C}), and construct the covariance matrix $\sigma_{ij}$ as defined in Eq.~\ref{cvdef}. By using this covariance matrix (Eq.~\ref{cvdef}) we can calculate the negativity using Eq.~\ref{negdef}.
\par

Lastly, we compute the complexities- COP and complexity from operator-state mapping. Just like the two-oscillator model COP for this case also given by Eq.~\ref{compare2}, with following identification, 
\begin{equation} \label{apr1}
\alpha=\sigma_P,~~ \tau^2=4\,  \sigma^2_{Q P}-1+4\,i\,\sigma_{Q P}, ~~{\rm Re }(\gamma)= 2\,\sigma_{Q},
\end{equation}

where $\sigma_Q,\sigma_P,\sigma_{Q P}$ are defined in Eq.~\ref{correlator}. Also in all the subsequent analysis we will set the reference frequency to be $\omega_0=1$ following \cite{PhysRevD.105.046011}. Last but not the least, to compute the circuit complexity using operator-state mapping, we start with the reduced density matrix of the oscillator as mentioned in Eq.~\ref{rhoSstructure}. As mentioned earlier we are mostly interested in $\hat\rho_s^{1/2}$ and the corresponding corresponding effective wavefunction will be of the form mentioned in Eq.~\ref{effectivepsi} and Eq.~\ref{effectivepsi1}. 
After that the complexity can be computed using Eq.~\ref{complexity}.

\subsubsection{Time Evolution of Linear Entropy, Negativity and Complexity}
\begin{figure}[htb!]
\begin{center}
\scalebox{1.2}{\includegraphics{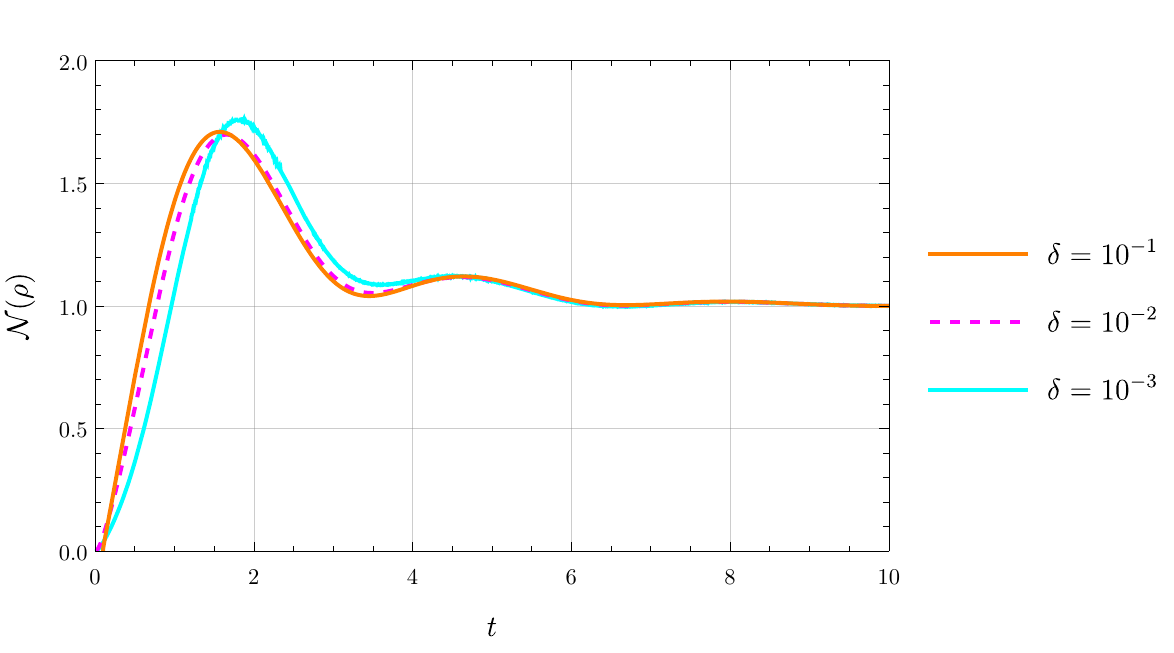} } 
\vspace{-0.1in}
\caption{Negativity vs time for different choices of $\delta$ (when $\omega_0 = 1$ and $\Gamma= 0.3$). The saturation value of $\mathcal{N}(\rho)$ is normalized to 1 in each plot.}
\label{fig:diffdelta}
\end{center}
\end{figure}

In this section we investigate in details  the saturation time scale for complexity, linear entropy and negativity. We have scanned the parameter space in detail. Below we pointwise summarize our results. 
\begin{itemize}
\item For both \textit{underdamped} and \textit{overdamped} cases we find that all four quantities saturate almost at the same time scale. The results are shown in Fig. \ref{fig:underdamped} (weak coupling, hence under damped) and Fig. \ref{fig:overdamped} (strong coupling, hence over damped).  
\begin{figure}[t!]
\begin{center}
\scalebox{1.2}{\includegraphics{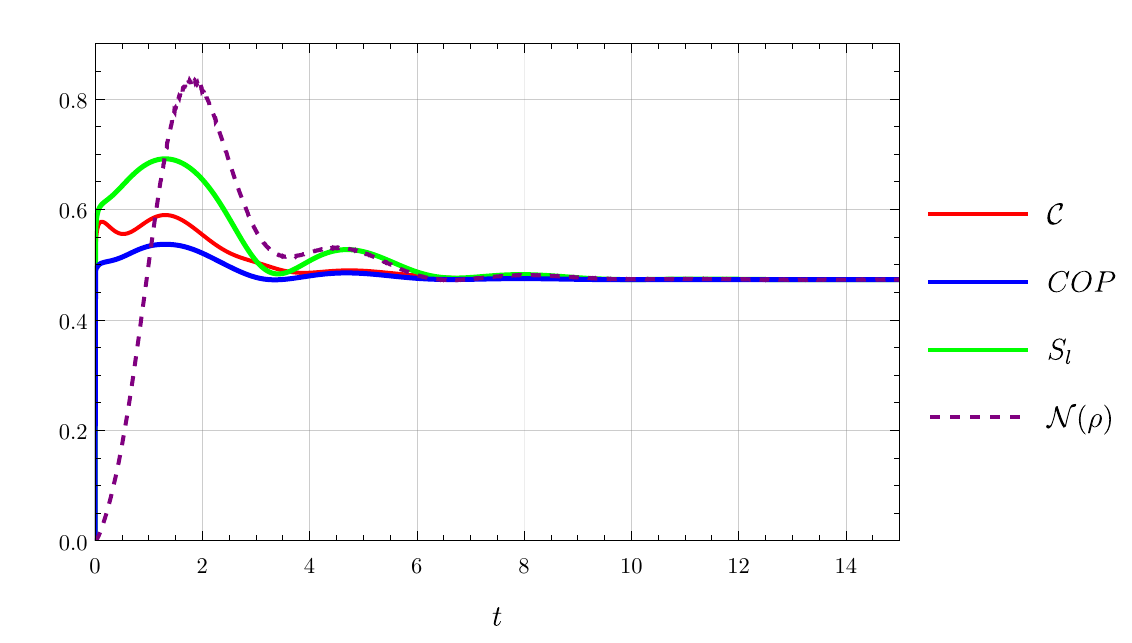} }
\vspace{-0.1in}
\caption{Complexity, Negativity and Linear Entropy vs time, by choosing $\omega_0 = 1$ and $\Gamma = 0.3$ in Caldeira Legget Model for Underdamped Oscillator.}
\label{fig:underdamped}
\end{center}
\end{figure}
\begin{figure}[b!]
\begin{center}
\scalebox{1.2}{\includegraphics{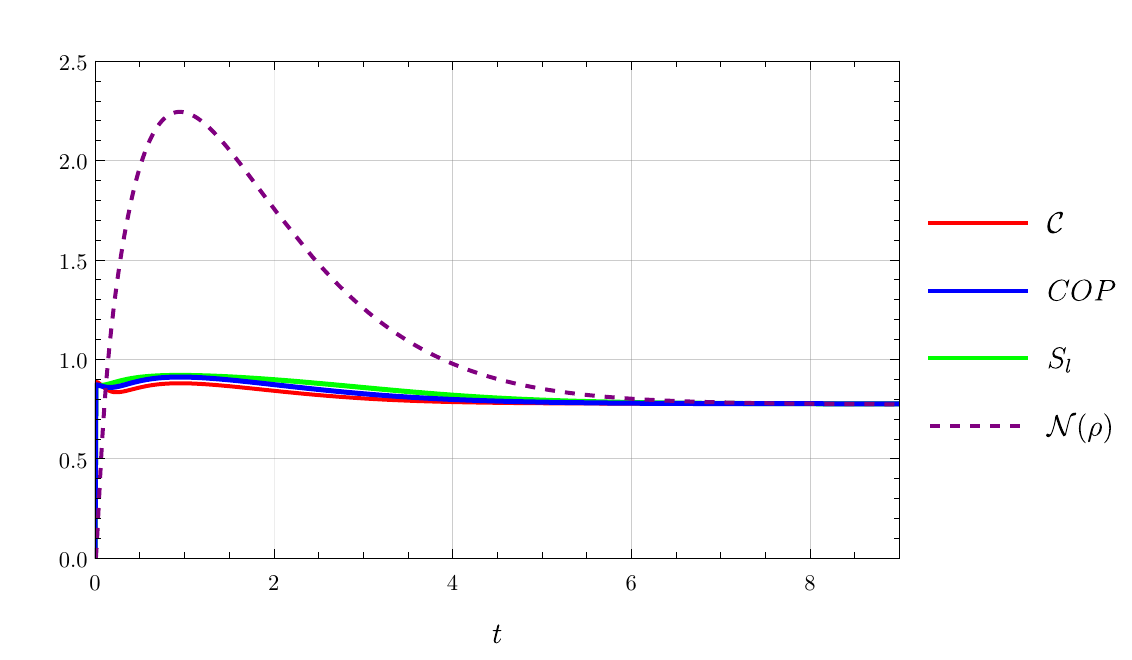} } 
\vspace{-0.1in}
\caption{Complexity, Negativity and Linear entropy vs time, by choosing $\Omega = 1;~ \Gamma = 1.2$ in Caldeira Legget Model for Overdamped Oscillators : }
\label{fig:overdamped}
\end{center}
\end{figure}
\item  Just like our previous model (the two-oscillator model), we find a similar behaviour for the time scales of linear entropy and complexity. When the system becomes completely \textit{mixed}--linear entropy saturates, both COP and complexity by operator state mapping is already saturated. The saturation time scale for complexity is smaller than the saturation time scale for linear entropy. So we can arrive at the same conclusion that, the complexity saturates when the system is maximally mixed. However, unlike the two oscillator model, the linear entropy and entanglement negativity have the same saturation time scale. This implies that for this model when the system is completely mixed then the quantum correlation/entanglement between the system and the bath also saturates. Therefore, entanglement negativity can detect when the system becomes mixed. 

\item Furthermore, our results imply that the complexity is more sensitive in detecting mixedness. To compute these quantities, we have computed various two-point correlators and used a cutoff to control various divergences that appear in these correlators. Details of the computation is mentioned in Appendix~\ref{C}. Moreover, as mentioned previously, we introduced a small length parameter $\delta$, and considered the bosonic bath fields at $x=\delta$ to compute the additional correlators. However, the saturation time scales do not depend on the choice of $\delta$ as shown in \cref{fig:diffdelta}, thereby making our claim about the saturation time scale universal.
\end{itemize}

\section{Discussion} \label{discussion}
In this paper, we considered two different open quantum systems 
and compared the saturation time scale for complexity, linear entropy and entanglement negativity. Our first model, two coupled oscillators, is basically a toy model for open quantum system and the entanglement negativity does not saturate for this model. However, for a more realistic Caldeira Leggett model, we found that the time scale when the entanglement negativity saturates is of the same order as the time scale when the system is maximally mixed, i.e. linear entropy also saturates. The quantum correlation between the system and the bath saturates when the system becomes maximally mixed. 

On top of this, we discovered that complexity saturates a bit earlier than the above saturation time scale of linear entropy for both the two-oscillator and Caldeira Leggett model, implying that complexity is already maximum when the system is completely mixed \footnote{The authors of \cite{delCampo:2019qdx}have explored the time evolution of purity and logarithmic negativity for conformal field theory in the presence of noise. They purified the system's state by using a Thermofield double type purification and then found the logarithmic negativity by computing Renyi-1/2. On the other hand, we have computed negativity for the full system, i.e. system and bath combined, as we wanted to explore the time evolution of quantum correlation between the system and the bath.This distinguishes our setup from  \cite{delCampo:2019qdx}.}. We have thoroughly scanned the parameter space and found that these saturation time scales are independent of the regularization scheme (Caldeira Leggett model). All of these point to the universal nature of our results.\par
It will be interesting to generalize our computation for multi-oscillator (more than two) systems. One can start with a Caldeira Leggett type of model, but the model the bath by considering the sum of many oscillators. Also, in this paper,  we have considered the Caldeira Leggett model with the Dirichlet boundary condition. It will be worthwhile to understand the effect of the boundary condition; hence it will be interesting to repeat this study for other boundary conditions, e.g. Neumann boundary condition. Note that, in this paper, we have computed the entanglement negativity for full density matrix, i.e. system and bath combined. It will be interesting to increase the size of the system and compute the negativity for the system after bipartisan it and tracing out the bath. This will generate a true mixed state corresponding to the system, and then it will be interesting to study how the evolution of negativity for this mixed state. Suppose the system initially has some quantum correlation. In that case, studying the time of evolution negativity will help us understand how the initial quantum correlation changes when there is an interaction with the bath. \par
Furthermore, it will be very interesting to extend this study to conformal field theory (CFT). Some studies regarding decoherence dynamics for CFT and time evolution of entanglement negativity and purity have been done here \cite{delCampo:2019qdx}. It will be good to again compare the saturation time scales of these quantities with the corresponding complexity. This will open up the possibility of connecting with gravity by making use of the AdS/CFT correspondence \footnote{In fact, a possible gravity dual of decoherence dynamics has been proposed in \cite{delCampo:2019qdx}. Perhaps one can utilize that to investigate the dynamics of complexity as well.}.\par
Last but not least, there are other notions of complexity, e.g. Krylov complexity \cite{Parker_2019}, which measures operator growth under time evolution. Perhaps it is interesting to study this notion of complexity for the system we considered in this paper and investigate (perhaps along the line of \cite{Bhattacharya:2022gbz}) whether it can tell us anything about the system being mixed or not. Apart from this, one perhaps also study other measures of correlation, e.g. out-of-time-ordered correlator \cite{Syzranov:2018ikh,Tuziemski:2019rnx,Chakrabarty:2018dov,haque2020squeezed}, the entanglement of purification \cite{Bhattacharyya:2018sbw,Bhattacharyya:2019tsi} for this kind of open systems and compare with the behaviour complexity, linear entropy and negativity that we have studied here. We hope to get back to some of these questions in a future publication.    

\section*{Acknowledgements}
We would like to thank Eugene Kim for collaboration at an earlier stage of this work.  A.B is supported by Start-Up Research Grant (SRG/2020/001380), Mathematical Research Impact Centric Support Grant (MTR/2021/000490) by the Department of Science and Technology Science and Engineering Research Board (India) and Relevant Research Project grant (202011BRE03RP06633-BRNS) by the Board Of Research In Nuclear Sciences (BRNS), Department of atomic Energy, India. 
\appendix
\numberwithin{equation}{section}
\section{Circuit Complexity} \label{d}
In this we will briefly review some aspects of computation of circuit complexity. We will use the Nielsen's method \cite{NL1,NL2,NL3}. Further details can be found in \cite{Jefferson}. The question we explore is the following: given a reference state and set of elementary quantum gates, what is the most efficient way to get a target state. Namely, how to find the most efficient quantum circuit that starts at that reference state (at $s=0$) and terminates at a target state ($s=1$)
\begin{equation}
    |\Psi_{s=1}\rangle = U (s=1) |\Psi_{s=0}\rangle,
\end{equation}
where $U$ is the unitary operator that takes the reference state to the target state.  We will represent the target sate $|\Psi_{s=1}\rangle$ as $|\Psi_T\rangle$ and the reference state $|\Psi_{s=0}\rangle$ as $|\Psi_R\rangle$ in the rest of the paper. We construct it from a continuous sequence of parametrized path ordered exponential of a  \textit{control} Hamiltonian operator 
\begin{equation}
U(s)= {\overleftarrow{\mathcal{P}}} \exp[- i \int_0^{s} \hspace{-0.1in} ds' H(s') ] \ .
\end{equation}
Here $s$ parametrizes a path in the space of the unitaries and given a set of elementary gates $M_I$, the  \textit{control} Hamiltonian can be written as
\begin{equation}
    H(s)= Y^{I}(s) M_{I}\, .
\end{equation} 
The coefficients $Y^I$ are the control functions that dictates which gate will act at a given value of the parameter. The control function is basically a tangent vector in the space of unitaries and satisfy the Schrodinger equation \footnote{From Eq.~\ref{invert} we have $Y^I(s) M_I= \frac{d U(s)}{ds}. U^{-1
}(s).$ $M_I's$ are taken to be the generators of some groups and can be normalized in a way that they satisfy $M_I M_J^T=\delta_{IJ}.$ Then finally we have, $ Y^{I}(s)= \textrm{Tr} \Big(\frac{d U(s)}{ds}. U^{-1
}(s). M_I^{T}\Big).$ Here $T$ denotes the transpose of matrix. For further details interested readers are referred to \cite{Jefferson}.}
\begin{equation} \label{invert}
\frac{d U(s)}{ds} = -i\, Y^I(s) M_I U(s)\,.
\end{equation}
Then we define a cost functional $\mathcal{F} (U, \dot U)$ as follows \footnote{The dot defines the derivative w.r.t $s$.}:
\begin{equation}
{\mathcal C}(U)= \int_0^1 \mathcal{F} (U, \dot U) ds\, .
\end{equation}
Minimizing this cost functional gives us the optimal circuit. There are different choices for the cost functional \cite{Jefferson}. In this paper we will consider
\begin{equation}
\mathcal{F}_2 (U, Y)  = \sqrt{\sum_I (Y^I)^2}\, .
\label{quadCost}
\end{equation}
In this paper, the target wave function $\psi_T( x_{1}, x_{2} ) = \langle x_1, x_2 | \psi_T \rangle$ takes the following form,
 takes the following form
\begin{equation}
 \psi_T( x_{1}, x_{2} ) = {\cal N}(t)
     \exp \left ( - \frac{1}{2} \left[ \Omega_{1}(t) x_{1}^2 + \Omega_{2}(t) x_{2}^2 
                    - 2\,\kappa(t) x_1 x_2 \right]  \right )  \ ,
\label{wavefunction1}
\end{equation} 
where $\Omega_{1}(t)$, $\Omega_{2}(t) \in {\bf C}$ 
(with ${\rm Re}[\Omega_{1}(t)]$, ${\rm Re}[\Omega_{2}(t)] > 0$) 
and $\kappa(t)$, ${\cal N}(t) \in {\bf R}$. For the reference state $\psi_{R}(x_1,x_2)$ $\Omega_1(t=0)=\Omega_2(t=0)=1$ and $\kappa(t=0)=0\,.$
Following \cite{me1}, we will take the elementary gates $M_I$ as the generators of the $GL (2, C)$ group. For details interested readers are referred to \cite{me1}. In all the cases, the complexity takes the form  \cite{me1} (due to the structure of the wave functions)
\begin{equation}
\mathcal{C} =\frac{1}{2} \sqrt{ \sum_{\alpha = 1,2} \left (
     \ln \left (| \hat \Omega_{\alpha}| \right)^2 
  + \tan^{-1} \left (-\frac{ \text{Im} ( \hat \Omega_{\alpha})}{\text{Re} ( \hat \Omega_{\alpha})} \right)^2 
  \right ) }  \ ,
  \label{complexity}
\end{equation}
 where $\alpha=1, 2$ and the normal mode frequencies are given by,
\begin{align}
\begin{split}
&\hat \Omega_1=\frac{1}{2}\Big(\Omega_1(t)+\Omega_2(t)+\sqrt{(\Omega_1(t)-\Omega_2(t))^2+ 4\kappa(t)^2}\Big), \\&\hat \Omega_2=\frac{1}{2}\Big(\Omega_1(t)+\Omega_2(t)-\sqrt{(\Omega_1(t)-\Omega_2(t))^2+ 4\kappa(t)^2}\Big).
\end{split}
\end{align}
\section{Mode expansion for Caldeira Leggett Model and Periodic Boundary Conditions} \label{A}
Here we give some details of the mode expansion of the bath Hamiltonian \cite{PhysRevD.105.046011}. The Hamiltonian is given by, 
\begin{equation}
 H = \frac{1}{2} \int_0^L \hspace{-0.1in} dx
     \left[ \Pi^2 + (\partial_x \phi)^2 \right]  \ .
\end{equation}
The field $\phi$ satisfies the periodic boundary condition $\phi(x+L) = \phi(x)$.  
By introducing the mode expansions
\begin{equation}
 \phi(x) = \frac{1}{\sqrt{L}} \sum_{k} e^{ikx} 
 \frac{1}{\sqrt{2\Omega_k} } ( a^{\phantom \dagger}_{k} + a^{\dagger}_{-k} )
 \ \ , \ \ 
 \Pi(x) = -i \frac{1}{\sqrt{L} } \sum_{k} e^{ikx} 
 \sqrt{\frac{\Omega_k}{2} } ( a^{\phantom \dagger}_k - a^{\dagger}_{-k} )  \ ,
\end{equation}
where $k = (2\pi/L) n$ with $n \in {\bf Z}$, and $\Omega_k = |k|$, we can diagonalize the Hamiltonian
\begin{equation}
 H = \sum_{k} \Omega_k ( a^{\dagger}_n a^{\phantom \dagger}_n + 1/2 )
 \ .
\end{equation}
Then we introduce the dual field $\theta$ such that $\Pi = - \partial_x \theta$. The mode expansion for $\theta$ is given by
\begin{equation}
 \theta(x) = \frac{1}{\sqrt{L} } \sum_{k} e^{ikx} 
   \frac{\rm sgn(k) }{\sqrt{2 \Omega_k} } ( a^{\phantom \dagger}_k - a^{\dagger}_{-k} )  \ .
\end{equation} 
Finally, one introduces the chiral fields $\phi_R$ and $\phi_L$, such that
\begin{equation}
 \phi = \phi_R + \phi_L  \ \ , \ \ 
 \theta = \phi_R - \phi_L  \ ;
\end{equation}
using the mode expansions for $\phi$ and $\theta$, one obtains
\begin{equation}
 \phi_R(x) = \frac{1}{\sqrt{L}} \sum_{k>0}  \frac{1}{\sqrt{2\Omega_k} } 
  \left( e^{ikx} a^{\phantom \dagger}_{k} + e^{-ikx} a^{\dagger}_{k} \right)
 \ \ , \ \ 
 \phi_L(x) = \frac{1}{\sqrt{L} } \sum_{k>0} \sqrt{\frac{\Omega_k}{2} } 
  \left( e^{-ikx} a^{\phantom \dagger}_{-k} + e^{ikx} a^{\dagger}_{-k} \right)  \ .
\label{chiralmode}
\end{equation} 
We are ultimately interested in $L \rightarrow \infty$ --- 
Eq.~\ref{chiralmode} becomes 
\begin{equation}
 \phi_R(x) = \int_0^{\infty} \hspace{-0.1in} \frac{d\omega}{\sqrt{2\pi} }  \frac{1}{\sqrt{2\omega} } 
  \left( e^{i \omega x} a^{\phantom \dagger}_{\omega} + e^{-i \omega x} a^{\dagger}_{\omega} \right)
 \ \ , \ \ 
 \phi_L(x) =  \int_0^{\infty} \hspace{-0.1in} \frac{d\omega}{\sqrt{2\pi} }  \frac{1}{\sqrt{2\omega} }   
  \left( e^{-i \omega x} a^{\phantom \dagger}_{-\omega} + e^{i \omega x} a^{\dagger}_{-\omega} \right)  \ .
\nonumber
\end{equation} 
\section{Characteristic Function} \label{B}
We now spell out some details about the the characteristic function following \cite{PhysRevD.105.046011} to make this paper self-contained. By using the density matrix
\begin{equation} \label{densityCF}
    \rho ( Q, Q') = \mathcal{N} {\rm exp} \left \{ -\frac{1}{2} \left(\alpha Q^2 + \alpha^* Q'^2 -c Q Q'\right)   \right \},
\end{equation}
where $\alpha =(\alpha_1 + i \alpha_2) \in \mathbf{C}$ and $c \in \mathbf{R}$, one can find the $l$'th moment for an arbitrary operator $\mathcal{O}$ as follows
\begin{equation}
    \langle \mathcal{O}^l \rangle = {\rm tr} (\rho \,\mathcal{O}^l)\,.
\end{equation}
By defining a characteristic function as follows
\begin{equation}
    {\rm CF} (\epsilon, t) \equiv {\rm tr} (\rho\, e^{-i \epsilon \mathcal{O}}),
\end{equation}
where $\epsilon$ is a real parameter, we can find the moments of all orders
\begin{equation}
    \langle \mathcal{O}^l (t) \rangle = \frac{\partial^l}{\partial (i \epsilon)^l} {\rm CF}(\epsilon, t) \ \Big |_{\epsilon=0}\,.
\end{equation}
Now, a characteristic function for the position and momentum operators can be written in terms of creation and annihilation operators $a$ and $a^{\dagger}$ as follows:
\begin{eqnarray} 
 {\rm CF} (\epsilon, t) \equiv {\rm tr} (\rho e^{-i \eta ( \epsilon a +\epsilon^* a^{\dagger})}),
\end{eqnarray}
where $\epsilon \in \mathbb{C}$ and $\eta \in \mathbb{R}$. This is also known as the Wigner characteristic function.\par
The creation and annihilation operators are
\begin{equation} \label{eert}
    a=\frac{1}{\sqrt{2 \Omega}} (\Omega Q+ i P)\ , \ \  a^{\dagger}=\frac{1}{\sqrt{2 \Omega}} (\Omega Q -i P) .
\end{equation}
By using Eq.~\ref{eert} we can write
\begin{eqnarray} \label{CF3}
 {\rm CF} (\lambda, \mu) \equiv {\rm tr} (\rho \ e^{-i (\lambda Q +\mu P)} ),
\end{eqnarray}
where $\lambda= \eta (\epsilon+\epsilon^*) \sqrt{\Omega/2}$ and $\mu= i \eta (\epsilon - \epsilon^*)/ (\sqrt{2 \Omega})$. By using the commutation relation $( [Q, P]= i )$, the exponent in Eq.~\ref{CF3} can be expanded as 
\begin{eqnarray}
  e^{-i (\lambda Q +\mu P)} &=& e^{-i \lambda Q} e^{-i\mu P} e^{\frac{1}{2}i\lambda \mu} \nonumber \\ 
  &=& e^{-\frac{1}{2} i\mu P} e^{ \frac{1}{2}i\mu P} e^{-i \lambda Q} e^{- \frac{1}{2} i\mu P} e^{- \frac{1}{2} i\mu P} e^{\frac{1}{2}i\lambda \mu} \nonumber \\
  &=& e^{- \frac{1}{2} i\mu P} e^{ -i\lambda (Q+\frac{\mu}{2})} e^{ -\frac{1}{2} i\mu P} e^{ \frac{1}{2} i\lambda \mu} \nonumber \\
  &=& e^{ -\frac{1}{2} i\mu P} e^{-i\lambda Q} e^{ \frac{1}{2} i\mu P}\,.
\end{eqnarray}
 Now we will evaluate the trace in the position representation, where
 \begin{eqnarray}
  Q|Q'\rangle &=& Q'|Q'\rangle\,, \\
  \langle Q'|Q''\rangle &=&\delta (Q'-Q'')\,,\\
  \int_{-\infty}^{\infty} dQ'|Q'\rangle \langle Q'| & =& 1\,.
 \end{eqnarray}
 Furthermore,
 \begin{eqnarray}
 e^{ \frac{1}{2} i\mu P} |Q'\rangle &=& {\Big|} Q' - \frac{\mu}{2}{\Big \rangle}\,, \\
\langle Q'| \ e^{ \frac{1}{2} i\mu P}&=& {\Big \langle} Q' + \frac{\mu}{2}{\Big |}\,.;
 \end{eqnarray}
Using these we get from \ref{CF3}
\begin{eqnarray}
 {\rm CF} (\lambda, \mu) & = & {\rm tr} \  e^{ -\frac{1}{2} i\mu P} \rho \  e^{ -\frac{1}{2} i\mu P}  e^{-i\lambda Q} \\
 & =& \int  e^{-i\lambda Q'} {\Big \langle} Q' -\frac{\mu}{2} |\rho| Q'+\frac{\mu}{2} {\Big \rangle} dQ'\,.
\end{eqnarray}
Now plugging the density matrix (Eq. \ref {densityCF}) in the above expression and performing the gaussian integrals we get
\begin{equation}
     {\rm CF} (\lambda, \mu) = {\rm exp} \left\{ -\frac{A}{2} \lambda^2 - \frac{A'}{2} \mu^2 +B \mu \lambda \right\}\,,
\end{equation}
where the parameters $A, A'$ and $B$ are
\begin{eqnarray}
A &=& \frac{1}{2 (\alpha_1 -c)}\,, \\
A' &=& \frac{|\alpha|^2-c^2}{2 (\alpha_1 -c)}\,,\\
B &=& -\frac{ \alpha_2}{2 (\alpha_1-c)}.
\end{eqnarray}
These parameters are related to the correlators as follows: 
\begin{eqnarray}
 \langle Q Q \rangle &=& - \frac{\partial^2}{\partial \lambda^2} {\rm CF (\lambda, \mu)} {\Big |}_{\mu=\lambda=0} = A\,, \\
 \langle P P \rangle &=& - \frac{\partial^2}{\partial \mu^2} {\rm CF (\lambda, \mu)} {\Big |}_{\mu=\lambda=0} = A'\,, \\
\frac{1}{2} \langle QP+PQ \rangle &=& - \frac{\partial^2}{\partial \mu \partial \lambda} {\rm CF (\lambda, \mu)} {\Big |}_{\mu=\lambda=0} = B\,.
\end{eqnarray}
\section{Correlation Functions of Caldeira Leggett Model} \label{C}
We need compute the full system two-point correlators 
$\langle \hat{O}_{\alpha}(t_1) \hat{O}_{\beta}(t_2) \rangle$, 
{where $\hat{O}_{\alpha,\beta}$ = $Q$, $P$, $\phi$ or $\Pi$} for computing various quantities mentioned in the main text. In this regard the correlators relating $Q_p(t)$, $\phi_0(\delta,t)$ and $\Pi_0(\delta,t)$ plays an important role.  Following \cite{PhysRevD.105.046011}, we have 
\begin{equation}
 Q_P(t) = \int_0^{\infty} \hspace{-0.1in} \frac{d\omega}{\sqrt{2\pi} } 
  \frac{1}{\sqrt{2\omega} } \left[
     f(\omega) e^{-i\omega t} a^{\phantom \dagger}_{\omega} 
  + f(-\omega) e^{i\omega t} a^{\dagger}_{\omega} \right]  
 \ \ {\rm with} \ \ 
   f(\omega) = \frac{i\omega \,2 \sqrt{2\Gamma} }
                     {\omega^2 + i2 \Gamma \omega - \omega_0^2}  \ .
\end{equation}
Using that the bath is taken to be initially in its ground state, we obtain
\begin{equation}
 \langle Q_p(t_1) Q_p(t_2) \rangle = \frac{2\Gamma}{\pi} \int_0^{\infty} \hspace{-0.1in} d\omega~
 e^{-i\omega(t_1 - t_2) } \frac{\omega}{ (\omega^2 - \omega_0^2 )^2 + (2\Gamma \omega)^2 } \ .
\nonumber
\end{equation} \label{xxcorrelator}
Carrying out the integral(s), we obtain the following,  
\begin{eqnarray}
 \langle Q_p(t_1) Q_p(t_2) \rangle & = & I_1 + i I_2 
\end{eqnarray}
where
\begin{eqnarray} \label{redun}
  I_1 & = & - \frac{1}{4 \pi \sqrt{\Gamma^2-\omega_0^2}} \left ( 
    2 \left[ \sin (\tau \omega_+) {\rm Si} (\tau \omega_+) 
  - \sin (\tau \omega^*_+) {\rm Si} (\tau \omega^*_+) \right]  \right.
 \nonumber \\  & & \hspace{-0.0in}  \left. 
 + \cos (\tau \omega_+) \left[ {\rm Ci} (\tau \omega_+) + {\rm Ci} (- \tau \omega_+) \right] 
 -  \cos (\tau \omega^*_+) \left[ {\rm Ci} (\tau \omega^*_+) + {\rm Ci} (- \tau \omega^*_+) \right] 
   \right )  \ ,
 \nonumber \\
 I_2 & = &  \frac{1}{i (4 \sqrt{\Gamma^2-\omega_0^2})} \left( \sin (\omega_+ \tau) -\sin (\omega_+^* \tau)\right) \ .
\end{eqnarray}
In Eq.~\ref{redun}, $$\omega_+ =\sqrt{\omega_0^2 + \ 2\Gamma (-\Gamma +\sqrt{\Gamma^2- \omega_0^2}) }, \omega_+^* = \sqrt{\omega_0^2 -  2\Gamma (-\Gamma +\sqrt{\Gamma^2- \omega_0^2}) },$$
$\tau = t_1 - t_2$, and ${\rm Si}(z)$ (${\rm Ci}(z)$) is the sine-integral (cosine-integral) function \cite{nist}. 
We also compute the other correlators \cite{PhysRevD.105.046011}:
\begin{eqnarray}  \label{xpcorrelator} 
 \langle \dot Q_p(t_1) Q_p(t_2) \rangle & = & J_1 + i J_2,
 \end{eqnarray}
 where
 \begin{eqnarray}
 J_1 & = & - \frac{1}{4 \sqrt{2} \pi \sqrt{\Gamma^2-\omega_0^2}} \left\{ 
 2  \left[ \omega_+ \cos (\tau \omega_+) {\rm Si} (\tau \omega_+) 
  -  \omega^*_+ \cos (\tau \omega^*_+) {\rm Si} (\tau \omega^*_+) \right]  \right.
 \nonumber \\  & & \hspace{-0.0in}  \left. 
 -  \omega_+\sin (\tau \omega_+) \left[ {\rm Ci} (\tau \omega_+) + {\rm Ci} (- \tau \omega_+) \right] 
 +  \omega^*_+\sin (\tau \omega^*_+) \left[ {\rm Ci} (\tau \omega^*_+) + {\rm Ci} (- \tau \omega^*_+) \right] 
   \right\}  \ ,
 \nonumber \\
J_2 & = & \frac{1}{i (4 \sqrt{\Gamma^2-\omega_0^2})} \left( \omega_+ \cos (\omega_+ \tau) - \omega_+^* \cos (\omega_+^* \tau) \right).
\end{eqnarray}
Similarly we get
\begin{eqnarray}  \label{ppcorrelator} 
 \langle \dot Q_p(t_1) \dot Q_p(t_2) \rangle & = & K_1 + i K_2,
 \end{eqnarray}
 where
 \begin{eqnarray}
 K_1 & = & - \frac{1}{8 \pi \sqrt{\Gamma^2-\omega_0^2}} \left\{ 
 2  \left[ \omega_+^2 \sin (\tau \omega_+) {\rm Si} (\tau \omega_+) 
  -  \omega^{*2}_+ \sin (\tau \omega^*_+) {\rm Si} (\tau \omega^*_+) \right]  \right.
 \nonumber \\  & & \hspace{-0.0in}  \left. 
 + \omega_+^2 \cos (\tau \omega_+) \left[ {\rm Ci} (\tau \omega_+) + {\rm Ci} (- \tau \omega_+) \right] 
 -  \omega^{*2}_+\cos (\tau \omega^*_+) \left[ {\rm Ci} (\tau \omega^*_+) + {\rm Ci} (- \tau \omega^*_+) \right] 
   \right\}  \ ,
 \nonumber \\
K_2 & =& \frac{1}{i (4 \sqrt{\Gamma^2-\omega_0^2})} \left ( \omega_+^2 \sin (\omega_+ \tau) - \omega_+^{*2} \sin (\omega_+^* \tau) \right )\,.
\end{eqnarray}
Lastly, 
\begin{eqnarray}  
 \langle Q(t_1) Q_p(t_2) \rangle & = & 0
 \end{eqnarray}
and the correlators at time $t=0$:
\begin{equation} 
 \langle Q Q \rangle  =  \frac{1}{2 \Omega}, \ \ \ \langle \dot Q \dot Q \rangle  =  \frac{1}{2 \Omega}.
 \end{equation}
 Note that to carry out the computations discussed in the paper, we need equal time correlators as mentioned in Eq.~\ref{correlator}. But in the limit $t_1 \rightarrow t_2$, they are typically divergent. To regularize the integral, we have used a cutoff $\tau_c.$ \par
  The correlators with the (free) bosonic fields\footnote{as defined in Eqn. (\ref{bosdef})}, $\phi_0(\delta,t)$ and $\Pi_0(\delta,t)$, can be calculated to be:
 \begin{align}
 \langle\phi_0\,\phi_0 \rangle &= \int_0^\infty \frac{d\omega}{\pi\omega} \, \sin^2{(\omega\delta)} \\
 \langle\Pi_0\,\Pi_0 \rangle &= \int_0^\infty d\omega\, \frac{\omega}{\pi} \, \sin^2{(\omega\delta)} \\ 
 \langle\phi_0\,\Pi_0 + \Pi_0\,\phi_0 \rangle &= 0 \\
 \langle\phi_0\,Q_P + Q_P\,\phi_0\ \rangle &= \int_0^\infty \frac{d\omega}{\pi\omega} \, \sin{(\omega\delta)} \,{\rm Im}\left[f(\omega)\right]\\
 \langle\phi_0\,\dot{Q}_P + \dot{Q}_P\,\phi_0\ \rangle &= \int_0^\infty \frac{d\omega}{\pi} \, \sin{(\omega\delta)} \,{\rm Re}\left[f(\omega)\right]\\
 \langle\Pi_0\,Q_P + Q_P\,\Pi_0\ \rangle &= \int_0^\infty \frac{d\omega}{\pi} \, \sin{(\omega\delta)} \,{\rm Re}\left[f(\omega)\right]\\
 \langle\Pi_0\,\dot{Q}_P + \dot{Q}_P\,\Pi_0\ \rangle &= \int_0^\infty d\omega\,\frac{\omega}{\pi} \, \sin{(\omega\delta)} \,{\rm Im}\left[f(\omega)\right] \label{PiQp}
 \end{align}
However, the correlator in Eq. \ref{PiQp} is divergent. In order to regularize the integral, we look at the asymptotic series expansion of the integrand and indetify the divergent term. Then we subtract it from the original integrand to obtain a finite answer.



\bibliography{main}

\providecommand{\href}[2]{#2}\begingroup\raggedright\begin{thebibliography}{100}

\bibitem{horodecki2013fundamental}
M.~Horodecki and J.~Oppenheim,  {\em Fundamental limitations for quantum and
  nanoscale thermodynamics}, Nature communications {\bf 4} (2013), no.~1, 1--6.

\bibitem{Skrzypczyk_2014}
P.~Skrzypczyk, A.~J. Short and S.~Popescu,  {\em Work extraction and
  thermodynamics for individual quantum systems}, Nature Communications {\bf 5}
  (jun, 2014).

\bibitem{Lesgourgues:1996jc}
J.~Lesgourgues, D.~Polarski and A.~A. Starobinsky,  {\em {Quantum to classical
  transition of cosmological perturbations for nonvacuum initial states}},
  Nucl. Phys. B {\bf 497} (1997) 479--510
  [\href{http://www.arXiv.org/abs/gr-qc/9611019}{{\tt gr-qc/9611019}}].

\bibitem{PhysRevA.52.R2493}
P.~W. Shor,  {\em Scheme for reducing decoherence in quantum computer memory},
  Phys. Rev. A {\bf 52} (Oct, 1995) R2493--R2496.

\bibitem{Aharonov_2000}
D.~Aharonov,  {\em Quantum to classical phase transition in noisy quantum
  computers}, Physical Review A {\bf 62} (nov, 2000).

\bibitem{Habib:1998ai}
S.~Habib, K.~Shizume and W.~H. Zurek,  {\em {Decoherence, chaos, and the
  correspondence principle}}, Phys. Rev. Lett. {\bf 80} (1998) 4361--4365
  [\href{http://www.arXiv.org/abs/quant-ph/9803042}{{\tt quant-ph/9803042}}].

\bibitem{Schlosshauer:2019ewh}
M.~Schlosshauer,  {\em {Quantum Decoherence}}, Phys. Rept. {\bf 831} (2019)
  1--57 [\href{http://www.arXiv.org/abs/1911.06282}{{\tt 1911.06282}}].

\bibitem{Zurek:2003zz}
W.~H. Zurek,  {\em {Decoherence, einselection, and the quantum origins of the
  classical}}, Rev. Mod. Phys. {\bf 75} (2003) 715--775
  [\href{http://www.arXiv.org/abs/quant-ph/0105127}{{\tt quant-ph/0105127}}].

\bibitem{open}
I.~{Rotter} and J.~P. {Bird},  {\em {A review of progress in the physics of
  open quantum systems: theory and experiment}}, Reports on Progress in Physics
  {\bf 78} (Nov., 2015) 114001 [\href{http://www.arXiv.org/abs/1507.08478}{{\tt
  1507.08478}}].

\bibitem{Peters_2004}
N.~A. Peters, T.-C. Wei and P.~G. Kwiat,  {\em Mixed-state sensitivity of
  several quantum-information benchmarks}, Physical Review A {\bf 70} (nov,
  2004).

\bibitem{Horodecki_2009}
R.~Horodecki, P.~Horodecki, M.~Horodecki and K.~Horodecki,  {\em Quantum
  entanglement}, Reviews of Modern Physics {\bf 81} (jun, 2009) 865--942.

\bibitem{Peres:1996dw}
A.~Peres,  {\em {Separability criterion for density matrices}}, Phys. Rev.
  Lett. {\bf 77} (1996) 1413--1415
  [\href{http://www.arXiv.org/abs/quant-ph/9604005}{{\tt quant-ph/9604005}}].

\bibitem{Horodecki:1996nc}
M.~Horodecki, P.~Horodecki and R.~Horodecki,  {\em {On the necessary and
  sufficient conditions for separability of mixed quantum states}}, Phys. Lett.
  A {\bf 223} (1996) 1 [\href{http://www.arXiv.org/abs/quant-ph/9605038}{{\tt
  quant-ph/9605038}}].

\bibitem{Eisert:1998pz}
J.~Eisert and M.~B. Plenio,  {\em {A Comparison of entanglement measures}}, J.
  Mod. Opt. {\bf 46} (1999) 145--154
  [\href{http://www.arXiv.org/abs/quant-ph/9807034}{{\tt quant-ph/9807034}}].

\bibitem{Vidal:2002zz}
G.~Vidal and R.~F. Werner,  {\em {Computable measure of entanglement}}, Phys.
  Rev. A {\bf 65} (2002) 032314
  [\href{http://www.arXiv.org/abs/quant-ph/0102117}{{\tt quant-ph/0102117}}].

\bibitem{Plenio:2005cwa}
M.~B. Plenio,  {\em {Logarithmic Negativity: A Full Entanglement Monotone That
  is not Convex}}, Phys. Rev. Lett. {\bf 95} (2005), no.~9, 090503
  [\href{http://www.arXiv.org/abs/quant-ph/0505071}{{\tt quant-ph/0505071}}].

\bibitem{Verstraete_2001}
F.~Verstraete, K.~Audenaert, J.~Dehaene and B.~D. Moor,  {\em A comparison of
  the entanglement measures negativity and concurrence}, Journal of Physics A:
  Mathematical and General {\bf 34} (nov, 2001) 10327--10332.

\bibitem{PhysRevA.62.022310}
S.~Ishizaka and T.~Hiroshima,  {\em Maximally entangled mixed states under
  nonlocal unitary operations in two qubits}, Phys. Rev. A {\bf 62} (Jul, 2000)
  022310.

\bibitem{PhysRevA.67.022110}
T.-C. Wei, K.~Nemoto, P.~M. Goldbart, P.~G. Kwiat, W.~J. Munro and
  F.~Verstraete,  {\em Maximal entanglement versus entropy for mixed quantum
  states}, Phys. Rev. A {\bf 67} (Feb, 2003) 022110.

\bibitem{PhysRevA.64.012316}
F.~Verstraete, K.~Audenaert and B.~De~Moor,  {\em Maximally entangled mixed
  states of two qubits}, Phys. Rev. A {\bf 64} (Jun, 2001) 012316.

\bibitem{PhysRevLett.92.087901}
G.~Adesso, A.~Serafini and F.~Illuminati,  {\em Determination of Continuous
  Variable Entanglement by Purity Measurements}, Phys. Rev. Lett. {\bf 92}
  (Feb, 2004) 087901.

\bibitem{Adesso:2004hs}
G.~Adesso, A.~Serafini and F.~Illuminati,  {\em {Extremal entanglement and
  mixedness in continuous variable systems}}, Phys. Rev. A {\bf 70} (2004)
  022318 [\href{http://www.arXiv.org/abs/quant-ph/0402124}{{\tt
  quant-ph/0402124}}].

\bibitem{Benatti:2006pw}
F.~Benatti and R.~Floreanini,  {\em {Entangling oscillators through environment
  noise}}, J. Phys. A {\bf 39} (2006) 2689--2700
  [\href{http://www.arXiv.org/abs/quant-ph/0602045}{{\tt quant-ph/0602045}}].

\bibitem{Singh_2015}
U.~Singh, M.~N. Bera, H.~S. Dhar and A.~K. Pati,  {\em Maximally coherent mixed
  states: Complementarity between maximal coherence and mixedness}, Physical
  Review A {\bf 91} (may, 2015).

\bibitem{delCampo:2019qdx}
A.~Del~Campo and T.~Takayanagi,  {\em {Decoherence in Conformal Field Theory}},
  JHEP {\bf 02} (2020) 170 [\href{http://www.arXiv.org/abs/1911.07861}{{\tt
  1911.07861}}].

\bibitem{Susskind:2014moa}
L.~Susskind,  {\em {Entanglement is not enough}}, Fortsch. Phys. {\bf 64}
  (2016) 49--71 [\href{http://www.arXiv.org/abs/1411.0690}{{\tt 1411.0690}}].

\bibitem{Susskind:2014rva}
L.~Susskind,  {\em {Computational Complexity and Black Hole Horizons}},
  Fortsch. Phys. {\bf 64} (2016) 24--43
  [\href{http://www.arXiv.org/abs/1403.5695}{{\tt 1403.5695}}], [Addendum:
  Fortsch.Phys. 64, 44--48 (2016)].

\bibitem{Brown:2015bva}
A.~R. Brown, D.~A. Roberts, L.~Susskind, B.~Swingle and Y.~Zhao,  {\em
  {Holographic Complexity Equals Bulk Action?}}, Phys. Rev. Lett. {\bf 116}
  (2016), no.~19, 191301 [\href{http://www.arXiv.org/abs/1509.07876}{{\tt
  1509.07876}}].

\bibitem{Brown:2015lvg}
A.~R. Brown, D.~A. Roberts, L.~Susskind, B.~Swingle and Y.~Zhao,  {\em
  {Complexity, action, and black holes}}, Phys. Rev. D {\bf 93} (2016), no.~8,
  086006 [\href{http://www.arXiv.org/abs/1512.04993}{{\tt 1512.04993}}].

\bibitem{Carmi:2016wjl}
D.~Carmi, R.~C. Myers and P.~Rath,  {\em {Comments on Holographic Complexity}},
  JHEP {\bf 03} (2017) 118 [\href{http://www.arXiv.org/abs/1612.00433}{{\tt
  1612.00433}}].

\bibitem{Jefferson}
R.~Jefferson and R.~C. Myers,  {\em {Circuit complexity in quantum field
  theory}}, JHEP {\bf 10} (2017) 107
[\href{http://www.arXiv.org/abs/1707.08570}{{\tt 1707.08570}}].

\bibitem{Chapman:2017rqy}
S.~Chapman, M.~P. Heller, H.~Marrochio and F.~Pastawski,  {\em {Toward a
  Definition of Complexity for Quantum Field Theory States}}, Phys. Rev. Lett.
  {\bf 120} (2018), no.~12, 121602
  [\href{http://www.arXiv.org/abs/1707.08582}{{\tt 1707.08582}}].

\bibitem{Bhattacharyya:2018wym}
A.~Bhattacharyya, P.~Caputa, S.~R. Das, N.~Kundu, M.~Miyaji and T.~Takayanagi,
  {\em {Path-Integral Complexity for Perturbed CFTs}}, JHEP {\bf 07} (2018) 086
  [\href{http://www.arXiv.org/abs/1804.01999}{{\tt 1804.01999}}].

\bibitem{Caputa:2017yrh}
P.~Caputa, N.~Kundu, M.~Miyaji, T.~Takayanagi and K.~Watanabe,  {\em {Liouville
  Action as Path-Integral Complexity: From Continuous Tensor Networks to
  AdS/CFT}}, JHEP {\bf 11} (2017) 097
  [\href{http://www.arXiv.org/abs/1706.07056}{{\tt 1706.07056}}].

\bibitem{me1}
T.~Ali, A.~Bhattacharyya, S.~Shajidul~Haque, E.~H. Kim and N.~Moynihan,  {\em
  {Time Evolution of Complexity: A Critique of Three Methods}}, JHEP {\bf 04}
  (2019) 087 [\href{http://www.arXiv.org/abs/1810.02734}{{\tt 1810.02734}}].

\bibitem{Bhattacharyya:2018bbv}
A.~Bhattacharyya, A.~Shekar and A.~Sinha,  {\em {Circuit complexity in
  interacting QFTs and RG flows}}, JHEP {\bf 10} (2018) 140
  [\href{http://www.arXiv.org/abs/1808.03105}{{\tt 1808.03105}}].

\bibitem{Hackl:2018ptj}
L.~Hackl and R.~C. Myers,  {\em {Circuit complexity for free fermions}}, JHEP
  {\bf 07} (2018) 139 [\href{http://www.arXiv.org/abs/1803.10638}{{\tt
  1803.10638}}].

\bibitem{Khan:2018rzm}
R.~Khan, C.~Krishnan and S.~Sharma,  {\em {Circuit Complexity in Fermionic
  Field Theory}}, Phys. Rev. D {\bf 98} (2018), no.~12, 126001
  [\href{http://www.arXiv.org/abs/1801.07620}{{\tt 1801.07620}}].

\bibitem{Camargo:2018eof}
H.~A. Camargo, P.~Caputa, D.~Das, M.~P. Heller and R.~Jefferson,  {\em
  {Complexity as a novel probe of quantum quenches: universal scalings and
  purifications}}, Phys. Rev. Lett. {\bf 122} (2019), no.~8, 081601
  [\href{http://www.arXiv.org/abs/1807.07075}{{\tt 1807.07075}}].

\bibitem{Ali:2018aon}
T.~Ali, A.~Bhattacharyya, S.~Shajidul~Haque, E.~H. Kim and N.~Moynihan,  {\em
  {Post-Quench Evolution of Complexity and Entanglement in a Topological
  System}}, Phys. Lett. B {\bf 811} (2020) 135919
  [\href{http://www.arXiv.org/abs/1811.05985}{{\tt 1811.05985}}].

\bibitem{Caputa:2018kdj}
P.~Caputa and J.~M. Magan,  {\em {Quantum Computation as Gravity}}, Phys. Rev.
  Lett. {\bf 122} (2019), no.~23, 231302
  [\href{http://www.arXiv.org/abs/1807.04422}{{\tt 1807.04422}}].

\bibitem{Guo:2018kzl}
M.~Guo, J.~Hernandez, R.~C. Myers and S.-M. Ruan,  {\em {Circuit Complexity for
  Coherent States}}, JHEP {\bf 10} (2018) 011
  [\href{http://www.arXiv.org/abs/1807.07677}{{\tt 1807.07677}}].

\bibitem{Bhattacharyya:2019kvj}
A.~Bhattacharyya, P.~Nandy and A.~Sinha,  {\em {Renormalized Circuit
  Complexity}}, Phys. Rev. Lett. {\bf 124} (2020), no.~10, 101602
[\href{http://www.arXiv.org/abs/1907.08223}{{\tt 1907.08223}}].

\bibitem{Flory:2020eot}
M.~Flory and M.~P. Heller,  {\em {Geometry of Complexity in Conformal Field
  Theory}}, Phys. Rev. Res. {\bf 2} (2020), no.~4, 043438
  [\href{http://www.arXiv.org/abs/2005.02415}{{\tt 2005.02415}}].

\bibitem{Erdmenger:2020sup}
J.~Erdmenger, M.~Gerbershagen and A.-L. Weigel,  {\em {Complexity measures from
  geometric actions on Virasoro and Kac-Moody orbits}}, JHEP {\bf 11} (2020)
  003 [\href{http://www.arXiv.org/abs/2004.03619}{{\tt 2004.03619}}].

\bibitem{Ali:2019zcj}
T.~Ali, A.~Bhattacharyya, S.~S. Haque, E.~H. Kim, N.~Moynihan and J.~Murugan,
  {\em {Chaos and Complexity in Quantum Mechanics}}, Phys. Rev. D {\bf 101}
  (2020), no.~2, 026021 [\href{http://www.arXiv.org/abs/1905.13534}{{\tt
  1905.13534}}].

\bibitem{Bhattacharyya:2019txx}
A.~Bhattacharyya, W.~Chemissany, S.~Shajidul~Haque and B.~Yan,  {\em {Towards
  the web of quantum chaos diagnostics}}, Eur. Phys. J. C {\bf 82} (2022),
  no.~1, 87 [\href{http://www.arXiv.org/abs/1909.01894}{{\tt 1909.01894}}].

\bibitem{cosmology1}
A.~Bhattacharyya, S.~Das, S.~Shajidul~Haque and B.~Underwood,  {\em
  {Cosmological Complexity}}, Phys. Rev. D {\bf 101} (2020), no.~10, 106020
  [\href{http://www.arXiv.org/abs/2001.08664}{{\tt 2001.08664}}].

\bibitem{cosmology2}
A.~Bhattacharyya, S.~Das, S.~S. Haque and B.~Underwood,  {\em {Rise of
  cosmological complexity: Saturation of growth and chaos}}, Phys. Rev. Res.
  {\bf 2} (2020), no.~3, 033273
  [\href{http://www.arXiv.org/abs/2005.10854}{{\tt 2005.10854}}].

\bibitem{DiGiulio:2020hlz}
G.~Di~Giulio and E.~Tonni,  {\em {Complexity of mixed Gaussian states from
  Fisher information geometry}},
  \href{http://www.arXiv.org/abs/2006.00921}{{\tt 2006.00921}}.

\bibitem{Caceres:2019pgf}
E.~Caceres, S.~Chapman, J.~D. Couch, J.~P. Hernandez, R.~C. Myers and S.-M.
  Ruan,  {\em {Complexity of Mixed States in QFT and Holography}}, JHEP {\bf
  03} (2020) 012 [\href{http://www.arXiv.org/abs/1909.10557}{{\tt
  1909.10557}}].

\bibitem{Bhattacharyya:2020art}
A.~Bhattacharyya, W.~Chemissany, S.~S. Haque, J.~Murugan and B.~Yan,  {\em {The
  Multi-faceted Inverted Harmonic Oscillator: Chaos and Complexity}}, SciPost
  Phys. Core {\bf 4} (2021) 002
  [\href{http://www.arXiv.org/abs/2007.01232}{{\tt 2007.01232}}].

\bibitem{Liu_2020}
F.~Liu, S.~Whitsitt, J.~B. Curtis, R.~Lundgren, P.~Titum, Z.-C. Yang, J.~R.
  Garrison and A.~V. Gorshkov,  {\em {Circuit complexity across a topological
  phase transition}}, Phys. Rev. Res. {\bf 2} (2020), no.~1, 013323
  [\href{http://www.arXiv.org/abs/1902.10720}{{\tt 1902.10720}}].

\bibitem{Susskind:2020gnl}
L.~Susskind and Y.~Zhao,  {\em {Complexity and Momentum}},
  \href{http://www.arXiv.org/abs/2006.03019}{{\tt 2006.03019}}.

\bibitem{Chen:2020nlj}
B.~Chen, B.~Czech and Z.-z. Wang,  {\em {Cutoff Dependence and Complexity of
  the CFT$_2$ Ground State}}, \href{http://www.arXiv.org/abs/2004.11377}{{\tt
  2004.11377}}.

\bibitem{Czech:2017ryf}
B.~Czech,  {\em {Einstein Equations from Varying Complexity}}, Phys. Rev. Lett.
  {\bf 120} (2018), no.~3, 031601
  [\href{http://www.arXiv.org/abs/1706.00965}{{\tt 1706.00965}}].

\bibitem{Chapman:2018hou}
S.~Chapman, J.~Eisert, L.~Hackl, M.~P. Heller, R.~Jefferson, H.~Marrochio and
  R.~C. Myers,  {\em {Complexity and entanglement for thermofield double
  states}}, SciPost Phys. {\bf 6} (2019), no.~3, 034
  [\href{http://www.arXiv.org/abs/1810.05151}{{\tt 1810.05151}}].

\bibitem{Chapman:2019clq}
S.~Chapman and H.~Z. Chen,  {\em {Complexity for Charged Thermofield Double
  States}}, \href{http://www.arXiv.org/abs/1910.07508}{{\tt 1910.07508}}.

\bibitem{Doroudiani:2019llj}
M.~Doroudiani, A.~Naseh and R.~Pirmoradian,  {\em {Complexity for Charged
  Thermofield Double States}}, JHEP {\bf 01} (2020) 120
  [\href{http://www.arXiv.org/abs/1910.08806}{{\tt 1910.08806}}].

\bibitem{Geng:2019yxo}
H.~Geng,  {\em {$T\bar{T}$ Deformation and the Complexity=Volume Conjecture}},
  Fortsch. Phys. {\bf 68} (2020), no.~7, 2000036
  [\href{http://www.arXiv.org/abs/1910.08082}{{\tt 1910.08082}}].

\bibitem{Guo:2020dsi}
M.~Guo, Z.-Y. Fan, J.~Jiang, X.~Liu and B.~Chen,  {\em {Circuit complexity for
  generalized coherent states in thermal field dynamics}}, Phys. Rev. {\bf
  D101} (2020), no.~12, 126007
[\href{http://www.arXiv.org/abs/2004.00344}{{\tt 2004.00344}}].

\bibitem{Haque:2021hyw}
S.~S. Haque, C.~Jana and B.~Underwood,  {\em {Operator complexity for quantum
  scalar fields and cosmological perturbations}}, Phys. Rev. D {\bf 106}
  (2022), no.~6, 063510 [\href{http://www.arXiv.org/abs/2110.08356}{{\tt
  2110.08356}}].

\bibitem{Haque:2021kdm}
S.~S. Haque, C.~Jana and B.~Underwood,  {\em {Saturation of thermal complexity
  of purification}}, JHEP {\bf 01} (2022) 159
  [\href{http://www.arXiv.org/abs/2107.08969}{{\tt 2107.08969}}].

\bibitem{Caputa:2022yju}
P.~Caputa, N.~Gupta, S.~S. Haque, S.~Liu, J.~Murugan and H.~J.~R. Van~Zyl,
  {\em {Spread Complexity and Topological Transitions in the Kitaev Chain}},
  \href{http://www.arXiv.org/abs/2208.06311}{{\tt 2208.06311}}.

\bibitem{Caputa:2022eye}
P.~Caputa and S.~Liu,  {\em {Quantum complexity and topological phases of
  matter}}, \href{http://www.arXiv.org/abs/2205.05688}{{\tt 2205.05688}}.

\bibitem{Couch:2021wsm}
J.~Couch, Y.~Fan and S.~Shashi,  {\em {Circuit Complexity in Topological
  Quantum Field Theory}}, \href{http://www.arXiv.org/abs/2108.13427}{{\tt
  2108.13427}}.

\bibitem{Erdmenger:2021wzc}
J.~Erdmenger, M.~Flory, M.~Gerbershagen, M.~P. Heller and A.-L. Weigel,  {\em
  {Exact Gravity Duals for Simple Quantum Circuits}},
  \href{http://www.arXiv.org/abs/2112.12158}{{\tt 2112.12158}}.

\bibitem{Chagnet:2021uvi}
N.~Chagnet, S.~Chapman, J.~de~Boer and C.~Zukowski,  {\em {Complexity for
  Conformal Field Theories in General Dimensions}}, Phys. Rev. Lett. {\bf 128}
  (2022), no.~5, 051601 [\href{http://www.arXiv.org/abs/2103.06920}{{\tt
  2103.06920}}].

\bibitem{Koch:2021tvp}
R.~d.~M. Koch, M.~Kim and H.~J.~R. Van~Zyl,  {\em {Complexity from spinning
  primaries}}, JHEP {\bf 12} (2021) 030
  [\href{http://www.arXiv.org/abs/2108.10669}{{\tt 2108.10669}}].

\bibitem{Bhattacharyya:2022ren}
A.~Bhattacharyya, G.~Katoch and S.~R. Roy,  {\em {Complexity of warped
  conformal field theory}}, \href{http://www.arXiv.org/abs/2202.09350}{{\tt
  2202.09350}}.

\bibitem{Banerjee:2022ime}
A.~Banerjee, A.~Bhattacharyya, P.~Drashni and S.~Pawar,  {\em {CFT to BMS:
  Complexity and OTOC}}, \href{http://www.arXiv.org/abs/2205.15338}{{\tt
  2205.15338}}.

\bibitem{Bhattacharya:2022wlp}
A.~Bhattacharya, A.~Bhattacharyya and S.~Maulik,  {\em {Pseudo complexity of
  purification for free scalar field theories}},
  \href{http://www.arXiv.org/abs/2209.00049}{{\tt 2209.00049}}.

\bibitem{Chapman:2021jbh}
S.~Chapman and G.~Policastro,  {\em {Quantum Computational Complexity -- From
  Quantum Information to Black Holes and Back}},
  \href{http://www.arXiv.org/abs/2110.14672}{{\tt 2110.14672}}.

\bibitem{Bhattacharyya:2021cwf}
A.~Bhattacharyya,  {\em {Circuit complexity and (some of) its applications}},
  Int. J. Mod. Phys. E {\bf 30} (2021), no.~07, 2130005.

\bibitem{NL1}
M.~A. Nielsen,  {\em {A geometric approach to quantum circuit lower bounds}},
  Science {\bf 311} (2006), no.~4, 92
[\href{http://www.arXiv.org/abs/0502070}{{\tt 0502070}}].

\bibitem{NL2}
M.~R. Nielsen, M.~A.and~Dowling, M.~Gu and A.~M. Doherty,  {\em {Quantum
  Computation as Geometry}}, Science {\bf 311} (2006), no.~4, 1133--1135
[\href{http://www.arXiv.org/abs/0603161}{{\tt 0603161}}].

\bibitem{NL3}
M.~R. Nielsen, M.~A.and~Dowling,  {\em {The geometry of quantum computation}},
  Science {\bf 311} (2006), no.~4, 1133--1135
[\href{http://www.arXiv.org/abs/0701004}{{\tt 0701004}}].

\bibitem{open2}
A.~Bhattacharyya, S.~S. Haque and E.~H. Kim,  {\em {Complexity from the reduced
  density matrix: a new diagnostic for chaos}}, JHEP {\bf 10} (2021) 028
  [\href{http://www.arXiv.org/abs/2011.04705}{{\tt 2011.04705}}].

\bibitem{PhysRevD.105.046011}
A.~Bhattacharyya, T.~Hanif, S.~S. Haque and M.~K. Rahman,  {\em {Complexity for
  an open quantum system}}, Phys. Rev. D {\bf 105} (2022), no.~4, 046011
  [\href{http://www.arXiv.org/abs/2112.03955}{{\tt 2112.03955}}].

\bibitem{bengtsson2017geometry}
I.~Bengtsson and K.~{\.Z}yczkowski, {\em Geometry of Quantum States: An
  Introduction to Quantum Entanglement}.
\newblock Cambridge University Press, 2017.

\bibitem{PhysRevLett.77.1413}
A.~Peres,  {\em Separability Criterion for Density Matrices}, Phys. Rev. Lett.
  {\bf 77} (Aug, 1996) 1413--1415.

\bibitem{PhysRevA.65.032314}
G.~Vidal and R.~F. Werner,  {\em Computable measure of entanglement}, Phys.
  Rev. A {\bf 65} (Feb, 2002) 032314.

\bibitem{doi:10.1080/09500349908231260}
J.~Eisert and M.~B. Plenio,  {\em A comparison of entanglement measures},
  Journal of Modern Optics {\bf 46} (1999), no.~1, 145--154
  [\href{http://www.arXiv.org/abs/quant-ph/9807034}{{\tt quant-ph/9807034}}].

\bibitem{10.5555/2011706.2011707}
M.~B. Plbnio and S.~Virmani,  {\em {An Introduction to Entanglement Measures}},
  Quantum Info. Comput. {\bf 7} (Jan, 2007) 1–51
  [\href{http://www.arXiv.org/abs/quant-ph/0504163}{{\tt quant-ph/0504163}}].

\bibitem{Calabrese:2012ew}
P.~Calabrese, J.~Cardy and E.~Tonni,  {\em {Entanglement negativity in quantum
  field theory}}, Phys. Rev. Lett. {\bf 109} (2012) 130502
  [\href{http://www.arXiv.org/abs/1206.3092}{{\tt 1206.3092}}].

\bibitem{Calabrese:2012nk}
P.~Calabrese, J.~Cardy and E.~Tonni,  {\em {Entanglement negativity in extended
  systems: A field theoretical approach}}, J. Stat. Mech. {\bf 1302} (2013)
  P02008 [\href{http://www.arXiv.org/abs/1210.5359}{{\tt 1210.5359}}].

\bibitem{Calabrese:2013mi}
P.~Calabrese, L.~Tagliacozzo and E.~Tonni,  {\em {Entanglement negativity in
  the critical Ising chain}}, J. Stat. Mech. {\bf 1305} (2013) P05002
  [\href{http://www.arXiv.org/abs/1302.1113}{{\tt 1302.1113}}].

\bibitem{Alba_2013}
V.~Alba,  {\em {Entanglement negativity and conformal field theory: a Monte
  Carlo study}}, J. Stat. Mech. {\bf 1305} (2013) P05013
  [\href{http://www.arXiv.org/abs/1302.1110}{{\tt 1302.1110}}].

\bibitem{Calabrese:2014yza}
P.~Calabrese, J.~Cardy and E.~Tonni,  {\em {Finite temperature entanglement
  negativity in conformal field theory}}, J. Phys. A {\bf 48} (2015), no.~1,
  015006 [\href{http://www.arXiv.org/abs/1408.3043}{{\tt 1408.3043}}].

\bibitem{PhysRevB.90.064401}
C.-M. Chung, V.~Alba, L.~Bonnes, P.~Chen and A.~M. L\"auchli,  {\em
  Entanglement negativity via the replica trick: A quantum Monte Carlo
  approach}, Phys. Rev. B {\bf 90} (Aug, 2014) 064401.

\bibitem{Ruggiero_2016}
P.~Ruggiero, V.~Alba and P.~Calabrese,  {\em {Entanglement negativity in random
  spin chains}}, Phys. Rev. B {\bf 94} (2016), no.~3, 035152
  [\href{http://www.arXiv.org/abs/1605.00674}{{\tt 1605.00674}}].

\bibitem{Ruggieroa_2016}
P.~Ruggiero, V.~Alba and P.~Calabrese,  {\em {Negativity spectrum of
  one-dimensional conformal field theories}}, Phys. Rev. B {\bf 94} (2016),
  no.~19, 195121 [\href{http://www.arXiv.org/abs/1607.02992}{{\tt
  1607.02992}}].

\bibitem{Blondeau_Fournier_2016}
O.~Blondeau-Fournier, O.~A. Castro-Alvaredo and B.~Doyon,  {\em {Universal
  scaling of the logarithmic negativity in massive quantum field theory}}, J.
  Phys. A {\bf 49} (2016), no.~12, 125401
  [\href{http://www.arXiv.org/abs/1508.04026}{{\tt 1508.04026}}].

\bibitem{Eisler_2014}
V.~Eisler and Z.~Zimbor{\'{a}}s,  {\em Entanglement negativity in the harmonic
  chain out of equilibrium}, New Journal of Physics {\bf 16} (dec, 2014)
  123020.

\bibitem{Coser_2014}
A.~Coser, E.~Tonni and P.~Calabrese,  {\em {Entanglement negativity after a
  global quantum quench}}, J. Stat. Mech. {\bf 1412} (2014), no.~12, P12017
  [\href{http://www.arXiv.org/abs/1410.0900}{{\tt 1410.0900}}].

\bibitem{Hoogeveen_2015}
M.~Hoogeveen and B.~Doyon,  {\em {Entanglement negativity and entropy in
  non-equilibrium conformal field theory}}, Nucl. Phys. B {\bf 898} (2015)
  78--112 [\href{http://www.arXiv.org/abs/1412.7568}{{\tt 1412.7568}}].

\bibitem{PhysRevB.92.075109}
X.~Wen, P.-Y. Chang and S.~Ryu,  {\em Entanglement negativity after a local
  quantum quench in conformal field theories}, Phys. Rev. B {\bf 92} (Aug,
  2015) 075109.

\bibitem{PhysRevA.88.042319}
C.~Castelnovo,  {\em Negativity and topological order in the toric code}, Phys.
  Rev. A {\bf 88} (Oct, 2013) 042319.

\bibitem{PhysRevA.88.042318}
Y.~A. Lee and G.~Vidal,  {\em Entanglement negativity and topological order},
  Phys. Rev. A {\bf 88} (Oct, 2013) 042318.

\bibitem{Wen_2016}
X.~Wen, P.-Y. Chang and S.~Ryu,  {\em {Topological entanglement negativity in
  Chern-Simons theories}}, JHEP {\bf 09} (2016) 012
  [\href{http://www.arXiv.org/abs/1606.04118}{{\tt 1606.04118}}].

\bibitem{Wena_2016}
X.~Wen, S.~Matsuura and S.~Ryu,  {\em {Edge theory approach to topological
  entanglement entropy, mutual information and entanglement negativity in
  Chern-Simons theories}}, Phys. Rev. B {\bf 93} (2016), no.~24, 245140
  [\href{http://www.arXiv.org/abs/1603.08534}{{\tt 1603.08534}}].

\bibitem{Kudler_Flam_2019}
J.~Kudler-Flam and S.~Ryu,  {\em {Entanglement negativity and minimal
  entanglement wedge cross sections in holographic theories}}, Phys. Rev. D
  {\bf 99} (2019), no.~10, 106014
  [\href{http://www.arXiv.org/abs/1808.00446}{{\tt 1808.00446}}].

\bibitem{MohammadiMozaffar:2017chk}
M.~R. Mohammadi~Mozaffar and A.~Mollabashi,  {\em {Logarithmic Negativity in
  Lifshitz Harmonic Models}}, J. Stat. Mech. {\bf 1805} (2018), no.~5, 053113
  [\href{http://www.arXiv.org/abs/1712.03731}{{\tt 1712.03731}}].

\bibitem{PhysRevA.72.032334}
G.~Adesso and F.~Illuminati,  {\em Gaussian measures of entanglement versus
  negativities: Ordering of two-mode Gaussian states}, Phys. Rev. A {\bf 72}
  (Sep, 2005) 032334.

\bibitem{PhysRevLett.84.2726}
R.~Simon,  {\em Peres-Horodecki Separability Criterion for Continuous Variable
  Systems}, Phys. Rev. Lett. {\bf 84} (Mar, 2000) 2726--2729.

\bibitem{Baumgratz_2014}
T.~Baumgratz, M.~Cramer and M.~Plenio,  {\em Quantifying Coherence}, Physical
  Review Letters {\bf 113} (sep, 2014).

\bibitem{Girolami_2014}
D.~Girolami,  {\em Observable Measure of Quantum Coherence in Finite
  Dimensional Systems}, Physical Review Letters {\bf 113} (oct, 2014).

\bibitem{Streltsov:2015xia}
A.~Streltsov, U.~Singh, H.~S. Dhar, M.~N. Bera and G.~Adesso,  {\em {Measuring
  Quantum Coherence with Entanglement}}, Phys. Rev. Lett. {\bf 115} (2015),
  no.~2, 020403 [\href{http://www.arXiv.org/abs/1502.05876}{{\tt 1502.05876}}].

\bibitem{Bhattacharyya:2018sbw}
A.~Bhattacharyya, T.~Takayanagi and K.~Umemoto,  {\em {Entanglement of
  Purification in Free Scalar Field Theories}}, JHEP {\bf 04} (2018) 132
  [\href{http://www.arXiv.org/abs/1802.09545}{{\tt 1802.09545}}].

\bibitem{Bhattacharyya:2019tsi}
A.~Bhattacharyya, A.~Jahn, T.~Takayanagi and K.~Umemoto,  {\em {Entanglement of
  Purification in Many Body Systems and Symmetry Breaking}}, Phys. Rev. Lett.
  {\bf 122} (2019), no.~20, 201601
  [\href{http://www.arXiv.org/abs/1902.02369}{{\tt 1902.02369}}].

\bibitem{JAMIOLKOWSKI1972275}
A.~Jamiołkowski,  {\em Linear transformations which preserve trace and
  positive semidefiniteness of operators}, Reports on Mathematical Physics {\bf
  3} (1972), no.~4, 275 -- 278.

\bibitem{CHOI1975285}
M.-D. Choi,  {\em Completely positive linear maps on complex matrices}, Linear
  Algebra and its Applications {\bf 10} (1975), no.~3, 285 -- 290.

\bibitem{PhysRevA.87.022310}
M.~Jiang, S.~Luo and S.~Fu,  {\em Channel-state duality}, Phys. Rev. A {\bf 87}
  (Feb, 2013) 022310.

\bibitem{Hosur:2015ylk}
P.~Hosur, X.-L. Qi, D.~A. Roberts and B.~Yoshida,  {\em {Chaos in quantum
  channels}}, JHEP {\bf 02} (2016) 004
  [\href{http://www.arXiv.org/abs/1511.04021}{{\tt 1511.04021}}].

\bibitem{CALDEIRA1983374}
A.~Caldeira and A.~Leggett,  {\em Quantum tunnelling in a dissipative system},
  Annals of Physics {\bf 149} (1983), no.~2, 374--456.

\bibitem{PhysRevLett.46.211}
A.~O. Caldeira and A.~J. Leggett,  {\em Influence of Dissipation on Quantum
  Tunneling in Macroscopic Systems}, Phys. Rev. Lett. {\bf 46} (Jan, 1981)
  211--214.

\bibitem{zurek0}
W.~G. Unruh and W.~H. Zurek,  {\em Reduction of a wave packet in quantum
  Brownian motion}, Phys. Rev. D {\bf 40} (Aug, 1989) 1071--1094.

\bibitem{buttiker}
M.~B\"uttiker,  {\em Scattering theory of current and intensity noise
  correlations in conductors and wave guides}, Phys. Rev. B {\bf 46} (Nov,
  1992) 12485--12507.

\bibitem{Parker_2019}
D.~E. Parker, X.~Cao, A.~Avdoshkin, T.~Scaffidi and E.~Altman,  {\em {A
  Universal Operator Growth Hypothesis}}, Phys. Rev. X {\bf 9} (2019), no.~4,
  041017 [\href{http://www.arXiv.org/abs/1812.08657}{{\tt 1812.08657}}].

\bibitem{Bhattacharya:2022gbz}
A.~Bhattacharya, P.~Nandy, P.~P. Nath and H.~Sahu,  {\em {Operator growth and
  Krylov construction in dissipative open quantum systems}},
  \href{http://www.arXiv.org/abs/2207.05347}{{\tt 2207.05347}}.

\bibitem{Syzranov:2018ikh}
S.~Syzranov, A.~Gorshkov and V.~Galitski,  {\em {Out-of-time-order correlators
  in finite open systems}}, Phys. Rev. B {\bf 97} (2018), no.~16, 161114
  [\href{http://www.arXiv.org/abs/1704.08442}{{\tt 1704.08442}}].

\bibitem{Tuziemski:2019rnx}
J.~Tuziemski,  {\em {Out-of-time-ordered correlation functions in open systems:
  A Feynman-Vernon influence functional approach}}, Phys. Rev. A {\bf 100}
  (2019), no.~6, 062106 [\href{http://www.arXiv.org/abs/1903.05025}{{\tt
  1903.05025}}].

\bibitem{Chakrabarty:2018dov}
B.~Chakrabarty, S.~Chaudhuri and R.~Loganayagam,  {\em {Out of Time Ordered
  Quantum Dissipation}}, JHEP {\bf 07} (2019) 102
  [\href{http://www.arXiv.org/abs/1811.01513}{{\tt 1811.01513}}].

\bibitem{haque2020squeezed}
S.~S. Haque and B.~Underwood,  {\em Squeezed out-of-time-order correlator and
  cosmology}, Physical Review D {\bf 103} (jan, 2021).

\bibitem{nist}
F.~Olver, D.~Lozier, R.~Boisvert and C.~Clark, {\em NIST Handbook of
  Mathematical Functions}.
\newblock 01, 2010.

\end{thebibliography}\endgroup
\bibliographystyle{utphysmodb}







\end{document}